\documentclass[10pt,aps,pra,superscriptaddress,showpacs,floatfix,nofootinbib]{revtex4-2}

\usepackage{graphicx}
\usepackage{bm}
\usepackage{epstopdf}
\usepackage{color}
\usepackage{amsmath}
\usepackage{amssymb}
\usepackage{amsfonts}
\usepackage{array}
\usepackage{yfonts}
\usepackage{physics}
\usepackage{eucal}
\usepackage{tikz}
\usepackage{braket}
\usepackage{subfig}

\begin{document}

\title{Four-wave mixing and secondary radiations generated by nonharmonic two-color filaments in air: Influence of the Kerr and plasma nonlinearities}

\author{V. Tamulien{\.e}}
\affiliation{Laser Research Center, Vilnius University, Saul{\.e}tekio 10, Vilnius LT-10223, Lithuania}

\author{P. David}
\affiliation{Centre Lasers Intenses et Applications, Université de Bordeaux-CNRS-CEA, 33405 Talence Cedex, France}

\author{V. Vai\v{c}aitis}
\affiliation{Laser Research Center, Vilnius University, Saul{\.e}tekio 10, Vilnius LT-10223, Lithuania}

\author{M. Rebarz}
\affiliation{The Extreme Light Infrastructure ERIC, ELI Beamlines Facility Za Radnic{\'i} 835, 25241 Doln{\'i} B\v{r}e\v{z}any, Czech Republic}

\author{S. J. Espinoza}
\affiliation{The Extreme Light Infrastructure ERIC, ELI Beamlines Facility Za Radnic{\'i} 835, 25241 Doln{\'i} B\v{r}e\v{z}any, Czech Republic}

\author{F. Catoire}
\affiliation{Centre Lasers Intenses et Applications, Université de Bordeaux-CNRS-CEA, 33405 Talence Cedex, France}

\author{L. Bergé}
\email{luc.berge@u-bordeaux.fr}
\affiliation{Centre Lasers Intenses et Applications, Université de Bordeaux-CNRS-CEA, 33405 Talence Cedex, France}

\date{\today}

\begin{abstract} 
Four-wave mixing (FWM) is an efficient source of light waves emitted at various frequencies, usually associated with third-order optical nonlinearities. Whereas attention has mostly been paid in the past to the generation of Stokes (e.g., visible) modes by mixing two nonharmonic frequencies in degenerate FWM, the present work aims to analyze the weaker components, i.e., the anti-Stokes (resp. mid-IR) radiation and cascaded satellites, and characterize their conversion efficiency and tunability. Here we report the production of tunable mid-infrared radiation around 3.3 $\mu$m delivered by two-color femtosecond filaments in air combining $\sim 800$ nm fundamental and $\sim 1.3$ $\mu$m seed waves. Although the Kerr response of air plays a key role in creating both visible and mid-IR radiations, we show that plasma nonlinearities also contribute by broadening and mixing the pump frequencies. Two experimental setups are exploited to focus, separately, on the production of FWM modes and cascaded satellites - called secondary radiations. A first filamentation setup employing different focal lengths for the two colors displays an ionization-induced broadening of the mid-IR radiation during the plasma stage. A second setup focusing the two pump components over equal propagation distances allows us to unveil the secondary radiations emerging at lower intensity levels when plasma occurs. We examine which player, among the plasma- or Kerr-induced FWM, is the most active in the conversion process. Numerical simulations based on a local-current model and a unidirectional solver highlight the role of the Kerr stage in amplifying the FWM radiations prior to the development of the weaker satellites in plasma regime. Their results, agreeing well with the experimental data, demonstrate that a broad visible emission is needed to trigger secondary radiations.
\end{abstract}

\maketitle

\section{Introduction}
\label{sec1}

Frequency conversion of light through nonlinear media \cite{Boyd,Agrawal2012} provides an efficient way to produce various electromagnetic radiations spanning from the terahertz (THz) to XUV range. In particular, sources of coherent mid-infrared (mid-IR) radiation, beyond the 3 $\mu$m range, are highly relevant for a wide range of applications, encompassing terahertz (THz) and atto-sciences \cite{DiChiara2012,Colosimo2008,Auguste2012,Clerici2013,Nguyen2019,Jang2019}, ultrafast IR spectroscopy \cite{Calabrese2012} and laser material processing \cite{Werner2019}. Though various methods are currently available to supply coherent mid-IR radiation, those based on solid-state emitters \cite{Ma2019,Mirov2018} or fiber/waveguide supercontinuum sources \cite{Yu2013,Liu2014} have limited output power and spectral tunability \cite{Yao2012}. To obtain powerful, wavelength-tunable ultrashort mid-IR pulses, optical parametric chirped pulse amplifiers (OPCPA) have to be used \cite{Dubietis2023}, but this technology is limited in low-wavelength tunability and requires specific nonlinear crystals guaranteeing severe phase-matching constraints, high optical nonlinearity and good thermal conductivity.\\

By contrast, gaseous media, including ambient air, have broad transparency regions in the visible and infrared spectral ranges, and they can easily be exploited for nonlinear optical frequency conversion. Nowadays, gases interacting with intense femtosecond laser sources can be used for various types of emissions, some of them, related to THz pulse and high-order harmonic generation, have been deeply investigated in the past \cite{Li2020,Kim2008,Vaicaitis2018LP}. Low-order harmonic generation triggered by nonlinear conversion mechanisms such as four-wave mixing \cite{Vaicaitis2000,ThebergePRL2006}, third harmonic and supercontinuum generation \cite{Akozbeck2002,Tamuliene2020,Vaicaitis2018APB} or plasma-driven Brunel radiations \cite{Brunel1990,David2025} are still being explored.\\

Among these conversion mechanisms four-wave mixing (FWM), associated with the generation of wavelengths from ultraviolet \cite{Fuji2007_2} to the mid-infrared range \cite{Fuji2007} has been proven a powerful method to produce coherent radiations in spectral regions where no laser source is available, by coupling nonharmonic pump pulses in gases nonlinearly. Upon mixing a major frequency  (e.g., fundamental) $\omega_1$ and a minor one $\omega_2$ (second wave) in incommensurate ratio, the partially degenerate FWM scheme indeed creates strong Stokes $(\omega_1+\omega_1-\omega_2)$ and weaker anti-Stokes $(\omega_2+\omega_2-\omega_1)$ components. While many studies in this field already addressed the production of visible wavelengths in focused propagation geometry or in the filamentation regime \cite{ThebergePRL2006,Berge2008,Vaicaitis2018}, fewer works have been devoted to the production of mid- and far-infrared radiations \cite{Theberge2010,Fan2022} for which the question of the dominant conversion mechanism among Kerr response and photoionization is still debatable. In addition to the FWM modes, lower-amplitude cascaded satellites \cite{Kim2009} created upon secondary mixings of the pump frequencies have been reported in photocurrent-driven Brunel-type radiations \cite{Brunel1990} without receiving any clear identification so far.\\

In this paper we report two experimental demonstrations: i -- The first one is dedicated to generating not only visible radiation \cite{ThebergePRL2006}, but above all an efficient and tunable mid-IR emission through four-wave difference frequency mixing of femtosecond laser pulses in air, by coupling the fundamental radiation (FW, $\omega_1$) of a femtosecond Ti:Sapphire $\sim \, 800$ nm laser and a tunable signal wave (SW, $\omega_2$) optimally centered at $\sim \, 1300$ nm from an optical parametric amplifier (OPA), both propagating in filamentation regime \cite{Berge2007}. Preliminary mid-IR tunability data and energy scaling for a similar configuration have been reported recently \cite{Vaicaitis2026OQE}. Here we exploit a two-lens experimental setup allowing us to optimize mid-IR generation and, by operating at different energy levels, to disentangle the action of third-order Kerr nonlinearities from their plasma counterpart. Although the Kerr (cubic) response of gas media such as air sustains the occurrence of FWM visible radiations in laser filamentation \cite{Theberge2010,ThebergePRL2006,Vaicaitis2018}, the plasma response has also been numerically identified as a potential player in the supercontinuum generation related to filament-driven visible radiation \cite{Berge2008,Kim2009}. We clear up the respective role of the Kerr and plasma source terms. We confirm the key role of the Kerr nonlinearity in creating and amplifying the FWM modes for pump pulse energies limited to a few mJ, while plasma broadens more their spectra. ii -- The second experimental setup, close to the former one but operating at lower pump energy levels, is dedicated to the measurements and analysis of secondary radiations, which we find located at the sum and differences of the FW and SW frequencies, i.e., $\omega_1\pm\omega_2$, and their multiples. This second series of experiment focusing both the fundamental and signal pulses at the same focal distance, either in shortly or loosely focused geometry, corroborates the generic existence of secondary radiations at lower spectral intensity, provided that the FWM-induced visible radiation is effective and broad enough.\\

For the above two setups, experimental features are confirmed by in-depth numerical simulations and simpler analytical models. We report that, although the Kerr stage is essential in the generation of FWM frequencies, attosecond-scaled ionization events contribute to enlarge the output spectrum. Secondary radiations need both a broadened visible radiation and the action of plasma nonlinearities.\\

The paper is organized as follows: Section \ref{sec2} presents the two experimental setups and main related results. With a two-color, two-lens setup operating in meter-range filamentation, emphasis is first given to the generated mid-IR radiation, its conversion efficiency with respect to the pump (FW and SW) input energy, and its potential tunability. Kerr-driven FWM spectra collected at low FW energies are compared with their plasma-driven counterparts triggered at high FW energies. Second, complementary experiments using equal focal distances for the two colors reveal that secondary radiations can occur, distinct from the broadened pump and FWM emissions, as original plasma markers during the nonlinear coupling between the FW and SW pulses. These secondary radiations are located near the frequency combinations $\omega_1 \pm \omega_2$ and their harmonics. Section \ref{sec3} theoretically and numerically examines the generation of FWM radiations by an air plasma for the same pump frequency arrangement compared with instantaneous and Raman-delayed Kerr nonlinearities \cite{Penano2003,Nguyen2017,Stathopulos2023}. On the basis of local current (LC) computations \cite{Kim2009,Babushkin2011}, the plasma response is evaluated with respect to the Kerr nonlinearities in the laser-to-mid-IR and visible conversion process when the FW pump pulse has enough energy to trigger photo-ionization while the SW wave is non-ionizing. These results are then revisited in the light of comprehensive unidirectional simulations \cite{Kolesik2002,Kolesik2004}, whose results confirm that the early Kerr stage is essential in the preliminary amplification of the FWM radiations while the following plasma stage may not necessarily increase their amplitude, but it contributes to broaden the FWM modes. It is also shown that an enlarged visible spectrum is necessary for the production of secondary radiations. This broadened visible radiation furthermore causes the occurrence of photocurrent-driven THz radiation through its mutual coupling with the propagating SW pulse that contributes to ionize air near focus. Analytical calculations justify that such secondary radiations do need supercontinuum generated by two colors ionizing air. Section \ref{sec4} concludes this work.

\section{Experimental setups and measurements}
\label{sec2}

Two experiments were performed, both dedicated to the mid-IR and visible radiations generated in the filamentation regime by two-colored pump fields with incommensurate (non-harmonic) frequency ratio. The first experiment led at ELI Beamlines facility (Czech Republic) used a two-lens setup that allowed us to optimize and characterize performances in mid-IR pulse generation. The second setup at Laser Resarch Center (LRC), Vilnius (Lithuania) exploited a similar arrangement of two colors but these were focused at the same propagation distance and were operated at lower pump energy. This second setup was devoted to the development of weaker secondary radiations measured at frequencies being multiple of $\omega_1 \pm \omega_2$. The input laser parameters together with the pump and generated frequencies of interest (both FWM and secondary radiations) are summarized in Table \ref{Table1}. \\

\begin{table}
\begin{center}
\begin{tabular}{|l|c|r|}
\hline
ELI laser parameters & Notation & Value \\
\hline
FW wavelength & $\lambda_{1}$ & 802 nm\\
SW wavelength & $\lambda_{2}$ & 1292 nm\\
FW energy & $E_{10}$ & 1 mJ \\
SW energy & $E_{20}$ & 0.11-0.13 mJ \\
FW duration (FWHM) & $\tau$ & 35 fs\\
FW beam diameter & $2r_{10}$ & 13.5 mm \\
SW beam diameter & $2r_{20}$ & 3.5 mm \\
FW focal length & $f_{1}$ & 96 cm \\
SW focal length & $f_{2}$ & 75 cm \\
\hline
\end{tabular}
\hspace{1cm}
\begin{tabular}{|c|c|c|}
\hline
Wave & $\nu$ (THz) & $\lambda$ (nm)\\
\hline
FW & 374 & 802\\
SW & 232 & 1292\\
2FW-SW & 516 & 581 \\
2SW-FW & 90 & 3333 \\
3SW & 696 & 431 \\
2SW+FW & 838 & 357 \\
FW-SW & 142 & 2112 \\
2FW-2SW & 284 & 1056 \\
FW+SW  &  606 & 495 \\
\hline
\end{tabular}
\vskip 0.5cm
\begin{tabular}{|c|c|c|}
\hline
LRC laser parameters & Notation & Value \\
\hline
FW wavelength & $\lambda_{10}$ & 790 nm\\
SW wavelength & $\lambda_{20}$ & 1306 nm\\
FW energy & $E_{10}$ & 260 $\mu$J \\
SW energy & $E_{20}$ & 40 $\mu$J \\
FW duration (FWHM) & $\tau$ & 35 fs\\
FW beam diameter & $2r_{10}$ & 10.9 mm \\
SW beam diameter & $2r_{20}$ & 7.4 mm \\
FW focal length & $f_{1}$ & 20 cm \\
SW focal length & $f_{2}$ & 20 cm \\
\hline
\end{tabular}
\hspace{1cm}
\begin{tabular}{|c|c|c|}
\hline
Wave & $\nu$ (THz) & $\lambda$ (nm)\\
\hline
FW & 379.7 & 790\\
SW & 229.7 & 1306\\
2FW-SW & 529.8 & 566.3 \\
2SW-FW & 79.7 & 3764 \\
3SW & 689 & 435.4 \\
2SW+FW & 839 & 357.6 \\
FW-SW & 150 & 2000 \\
2FW-2SW & 300 & 1000 \\
FW+SW  &  609.5 & 492 \\
\hline
\end{tabular}
\caption{Top. Left: Input laser parameters for the ELI experiment. Right: Corresponding pump and generated frequencies. Bottom. Left: Input laser parameters for the LRC Vilnius experiment in focused propagation geometry. Right: Corresponding pump and generated frequencies.}
\label{Table1}
\end{center}
\end{table}

\subsubsection{The ELI setup}

\begin{figure*}[ht]
  \centering \includegraphics[width=\textwidth]{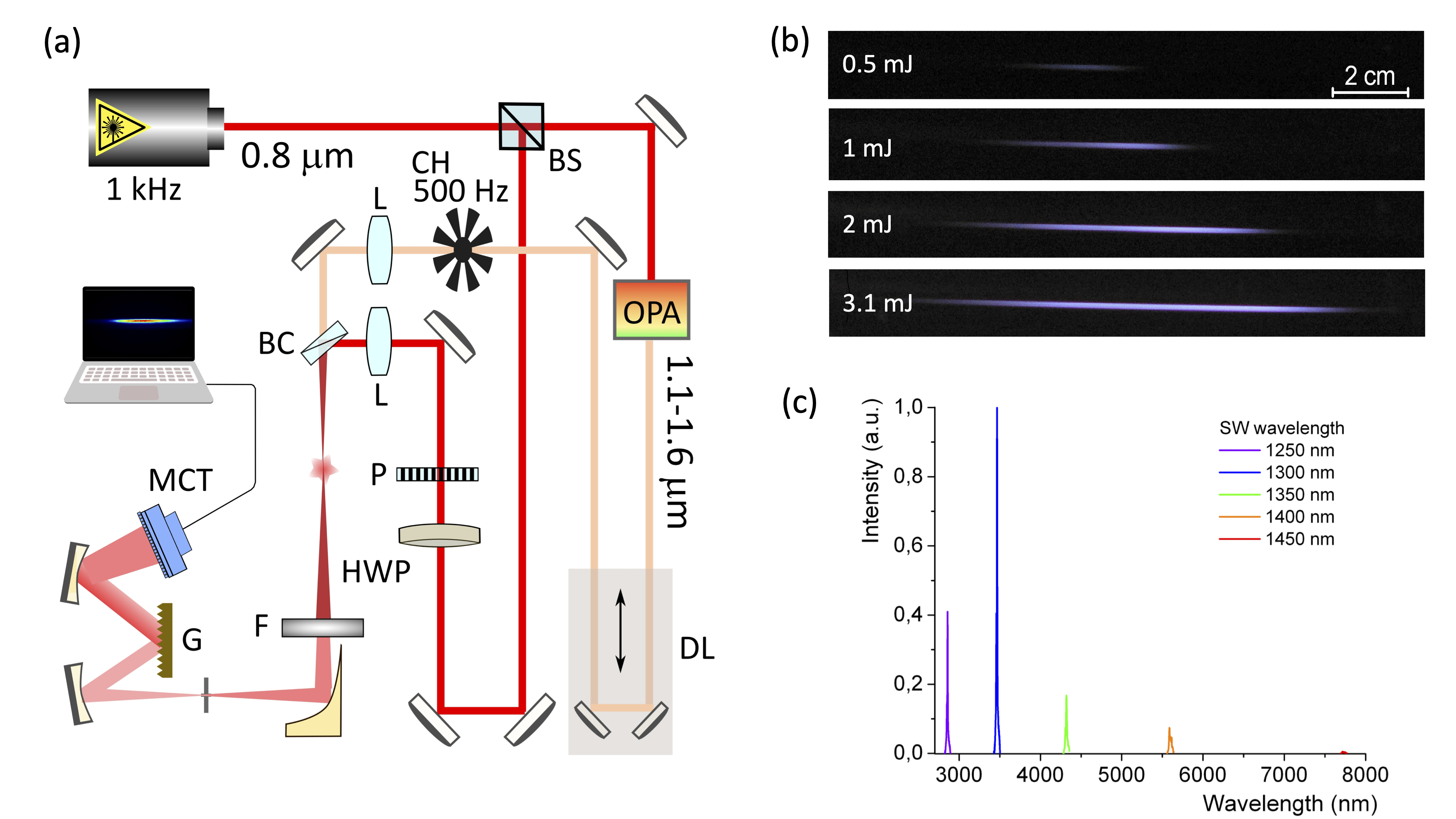} 
  \caption{(a) ELI experimental setup: BS – beam splitter, CH – chopper, DL – delay line, DM – dichroic mirror, F – IR long pass filter, GR – IR grating, L1, L2 – quartz lenses, MCT – HgCdTe detector, OPA – optical parametric amplifier, P – polarizer, WP – half wave plate. (b) Experimentally recorded plasma filaments, created by the focused FW beam at FW input energies of 0.5, 1, 2, and 3.1 mJ (top to bottom). (c) Spectra of the generated mid-IR radiation for SW wavelengths varying from 1.25 to 1.45 $\mu$m.}
\label{Fig1}
\end{figure*}

The ELI experimental setup is shown in Fig. \ref{Fig1} (see also \cite{Vaicaitis2026OQE}). An amplified Ti:Sapphire system (Coherent Astrella, 802 nm, 35 fs FWHM, 1 kHz, 6 mJ max.) provides the FW pump. Part of its energy (1.5 mJ approx.) seeds an OPA (TOPAS Prime) producing a 40-fs SW pulse tunable in 1.1–1.6 $\mu$m. The  SW wavelength  is  set  to 1292 nm -- where the OPA delivers its peak energy -- and is varied only for the tunability characterization shown in Fig. \ref{Fig1}(c). The two beams are made collinear by a dichroic mirror and focused by separate lenses L1 ($f_1 = 96$ cm, for the FW pulse) and L2 ($f_2 = 75$ cm, for the SW pulse), with L2 placed 21 cm downstream of L1, so that both beams share a common focus at $z = f_1$ from L1, with 1/e$^2$ beam diameters of 13.5 mm (FW) and 3.5 mm (SW). An optical delay line controls their relative timing; their polarizations are kept linear and parallel. Mid-IR radiation is filtered, focused by an off-axis parabolic mirror onto a Czerny–Turner spectrograph, and detected with a LN$_2$-cooled HgCdTe array (128 × 128, 1 kHz). The SW energy is kept at 0.11–0.13 mJ while the FW energy is varied between 0.022 and 3.1 mJ. The visible radiation is collimated with fused silica lenses and after proper optical filtering directed to the input of a fiber spectrometer or a laser powermeter. A 500-Hz-rated optical chopper inserted along the beam path stops the FW beam, enabling us to reduce noise and register the background-free radiation signal. All input parameters and the resulting frequency combinations are summarized in Table I (top).\\
  
When increasing the FW energy, plasma filaments created by the loosely focused FW beam could be observed from the fluorescence of the ionized air species [see Fig. \ref{Fig1}(b)]. At low FW energy $\sim 0.5$ mJ the filament length was about 2 cm, while it increased gradually to more than 14 cm at maximum FW energy (3.1 mJ). At the same time the brightness of the filament also increased with the FW energy, indicating an augmented ionization level. Note that the input SW energy was too low to ionize air and the length and brightness of the plasma filament did not depend on the fact if the SW beam was added to the FW beam or not. We checked that the output laser beam kept a smooth near-Gaussian intensity distribution after the beamsplitter and waveplate (see Fig. \ref{Fig1}). After passing the FW focusing lens and when generating a plasma filament, the FW pump spectrum was blueshifted and enlarged by $\Delta \lambda \sim 50$ nm around $800$ nm (not shown).\\

Figure \ref{Fig1}(c) displays evidence of the tunability of the generated mid-IR pulse. Maximum generation occurs for mid-IR frequency $\omega_{\rm mid-IR} = 2 \omega_{\rm SW} - \omega_{\rm FW} \equiv 2 \omega_{2} - \omega_{1}$ corresponding to the central wavelength $\lambda_{\rm mid-IR} = 2 \pi c/\omega_{\rm mid-IR} \approx 3.3\,\mu$m ($c$ denotes the speed of light in vacuum), which is reached for the SW wavelength $\sim$ $1.3\,\mu$m. Because the mid-IR radiation and the SW characteristics are strongly connected, a spectral analysis of the generated mid-IR radiation was performed within the SW wavelength tuning range between 1.25 and 1.45 $\mu$m, producing mid-IR wavelengths ranging from 2.86 to 7.73 $\mu$m. These data highlight the tunability of the generated mid-IR radiation. They, however, evidence a rapid decrease in the intensity of the generated mid-IR beam being strongly dependent on the SW wavelength and available SW power. Maximum SW pulse energy was reached at 1.29-1.3 $\mu$m and steadily decreased with the SW wavelength detuned from these values.\\

\begin{figure*}[ht]
  \centering \includegraphics[width=0.8\textwidth]{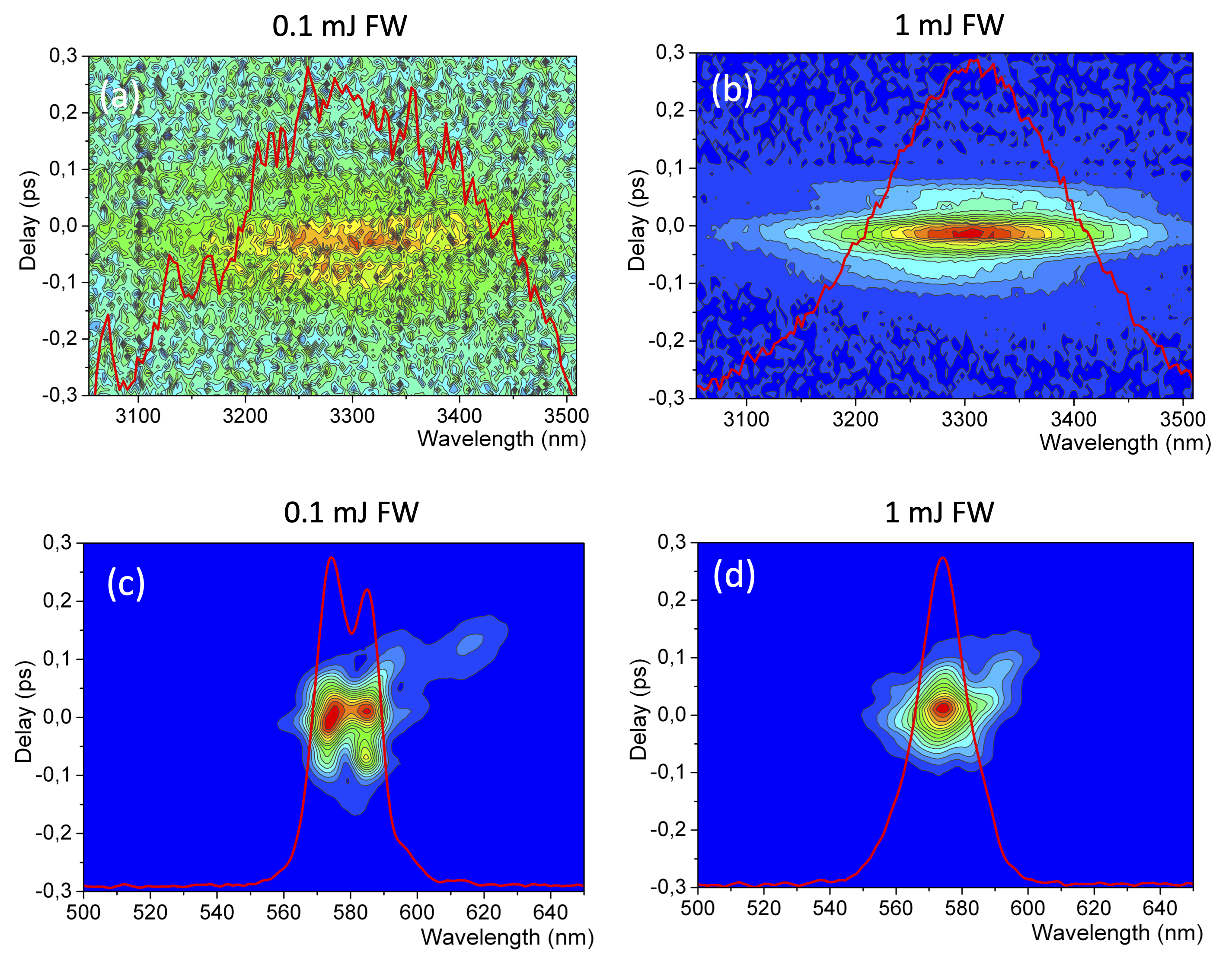}
  \caption{Spectra of (a,b) mid-IR radiation and (c,d) visible radiation as a function of time delay between the FW and SW pulses for (left) 0.1 mJ and (right) 1 mJ FW energy. Energy of the SW pulse is fixed at 0.13 mJ.}
\label{Fig2}
\end{figure*}

Figure \ref{Fig2} details the measured spectra of the mid-IR and visible radiations as a function of the time delay between the FW and SW pulses for two FW energies, 0.1 and 1 mJ, keeping the SW energy constant (0.13 mJ). As shown in Fig. \ref{Fig2}(a) for 0.1 mJ FW energy, coherent emission is generated in the mid-IR at the expected central wavelength $\lambda_{\rm mid-IR} \simeq$ 3300 nm (see \cite{Vaicaitis2026OQE} for complementary data), only taking place when the FW and SW pulses overlap in time. The simultaneously recorded visible radiation with frequency $\omega_{\rm vis} = 2 \omega_{\rm FW} - \omega_{\rm SW} \equiv 2 \omega_{1} - \omega_{2}$, centered around $\lambda_{\rm vis} \simeq 570-580$ nm in Fig. \ref{Fig2}(c), is generated again only when the FW and SW pulses overlap in time. Fluctuations can appear in the FWM generation process, here at low FW energy in the visible emission, which we attribute to the spectral broadening of the pump components due to cross and self-phase modulation. Figures \ref{Fig2}(b,d) show the same information on the FWM main spectral modes at 1 mJ FW energy, highlighting a stabilization and completion of both mid-IR and visible modes. Mid-IR radiation achieves a net broadening of $\sim 200$ nm in the presence of plasma.\\

\begin{figure*}[ht]
  \centering \includegraphics[width=\textwidth]{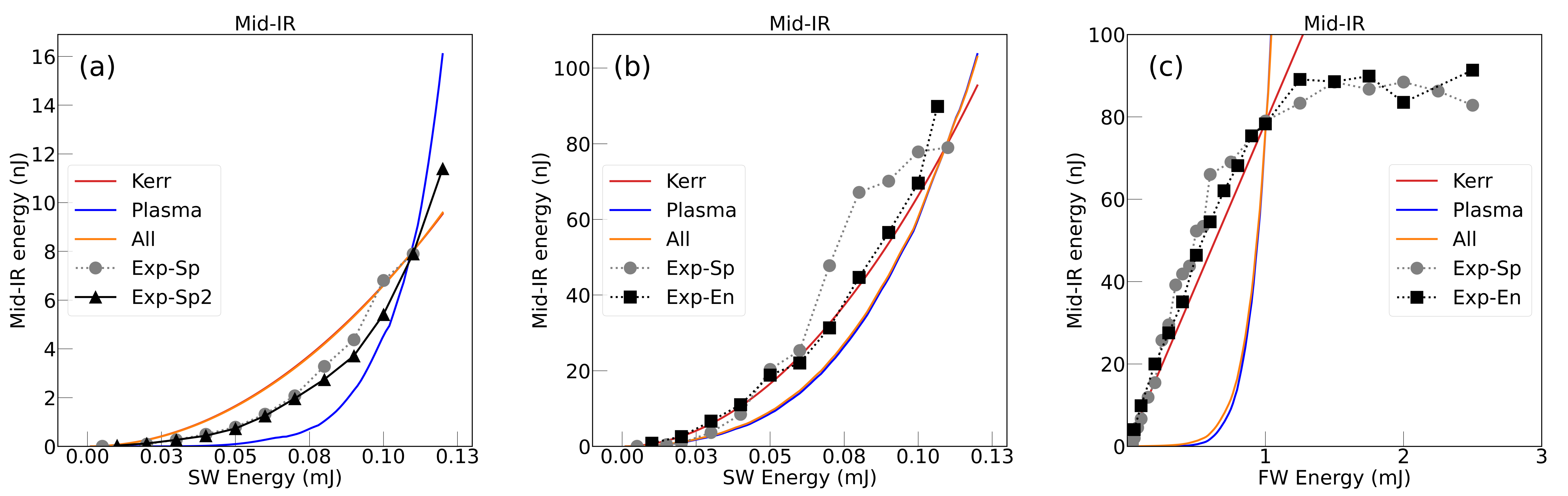}
  \caption{Energy of mid-IR radiation centered at 3300 nm as a function of (a) the SW energy for 0.1 mJ FW pulse (b) the SW energy for 1 mJ FW pulse and (c) the FW energy for 0.11 mJ SW pulse. In (a) data were extracted from two different spectral measurements (Exp-Sp and Exp-Sp2 - see legend and main text). In (b) and (c) spectral data (Sp) are compared with direct energy measurements (Exp-En - see legends and main text). In the colored curves showing results from the local-current model (see Sec.\ref{sec3A}), the energy calibration is made for the point corresponding to 0.11 mJ SW energy and 1 mJ FW energy delivering an energy yield of $\sim 79$ nJ in the mid-IR bandwidth.}
\label{Fig3}
\end{figure*}

\begin{figure*}[ht]
  \centering \includegraphics[width=\textwidth]{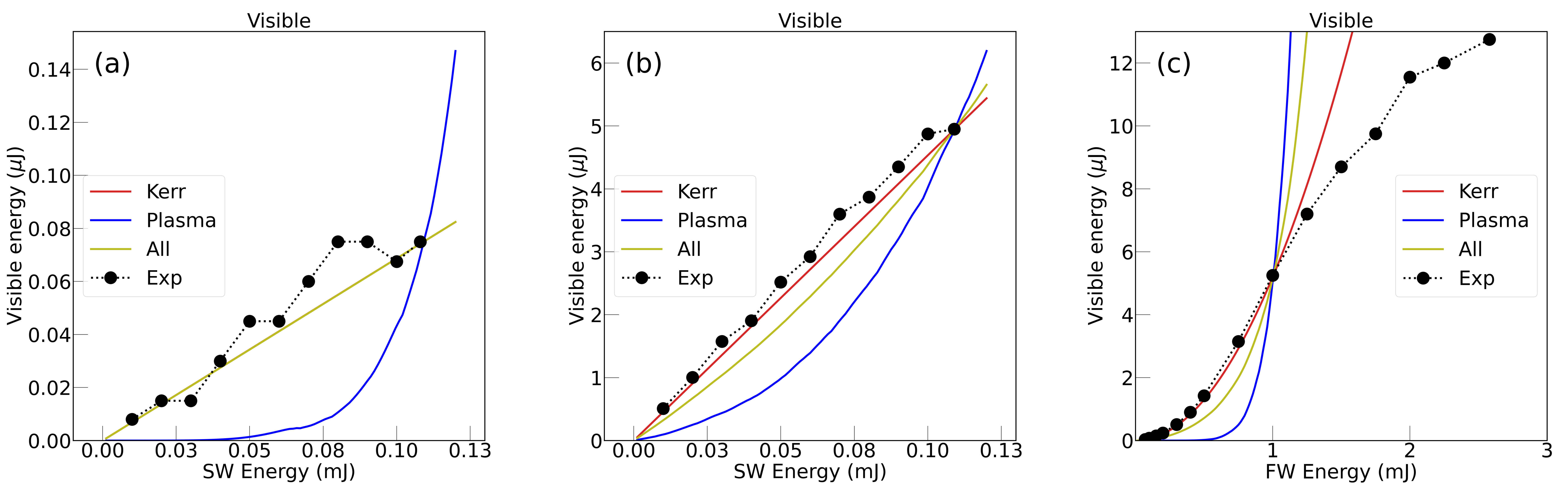}
  \caption{Energy of visible radiation centered at 580 nm as a function of (a) the SW energy for 0.1 mJ FW pulse (b) the SW energy for 1 mJ FW pulse and (c) the FW energy for 0.11 mJ SW pulse. Experimental data are extracted from energy measurements. The theoretical (LC) curves were calibrated at the point corresponding to 0.11 mJ SW energy and 1 mJ FW energy.}
\label{Fig4}
\end{figure*}

Figures \ref{Fig3}(a,b,c) next show the increase in the energy of the generated mid-IR beam as a function of the SW energy for a 0.1 mJ and 1 mJ FW pulses in Fig. \ref{Fig3}(a) and Fig. \ref{Fig3}(b), respectively, and as a function of the FW energy for a 0.11 mJ SW energy [Fig. \ref{Fig3}(c)]. In Fig. \ref{Fig3}(a) are plotted two sets of experimental data points, labelled Exp-Sp and Exp-Sp2, corresponding to spectral measurements taken at two different days along the experimental campaign to appreciate the reproducibility of the results. These data were collected at the output of the mid-IR spectrometer and were based on the varying amplitude and broadness of the measured spectra, requiring further energy calibration. One can observe that both data sets remain in relative close vicinity, so that only one of them (Sp) will henceforth be used in the next graphics. In the latter, another set of experimental data, labelled Exp-En, is instead shown as black dots linked by dotted lines: they consist in direct energy measurements collected using a calibrated energy-meter with optical filters placed in the beam path. Figures \ref{Fig4}(a-c) display similar information also based on energy measurements about the gain in the emitted visible energy. Typical laser-to-mid-IR conversion efficiency is evaluated at about $\sim 3 \times 10^{-5}$ leading to maximum mid-IR energy $\approx$ 80-90 nJ, while that of the generated visible light approaches $5 \times 10^{-3}$ with maximum energy of $\approx \,13\mu$J for a 2.5-mJ FW energy. In Figs. \ref{Fig3} and \ref{Fig4}, theoretical energy growths computed from intensity estimates based on the local current model discarding propagation effects (colored solid curves, see Sec. \ref{sec3}) are displayed for comparison. Their interpretation is detailed in Sec. \ref{sec3A}. 

\subsubsection{The LRC Vilnius setup}

Complementary experiments were performed at LRC, Vilnius, to acquire more and finer spectral data in frequency regions that were not scanned during the ELI experiment. Here, both pulse components were focused at the same linear distance $f = 20$ cm in focused propagation geometry or $f = 100$ cm in filamentation regime. Attention was paid to the bandwidths centered around $\omega_1 - \omega_2 \simeq 2 \pi \nu_{\rm S1}$, $2(\omega_1 - \omega_2) \simeq 2 \pi \nu_{\rm S2}$, and $\omega_1 + \omega_2 \simeq 2 \pi \nu_{\rm S3}$ related to new wavelengths associated with secondary radiations. Input laser parameters, pump and related frequency combinations are detailed in Table \ref{Table1} (bottom).\\

With the $f = 20$ cm setup, parallel, linearly polarized FW and SW pump pulses with equal FWHM duration $\sim 35$ fs, 1/e$^2$ beam diameters of 10.9 mm and 7.4 mm, and maximum energies of 0.45 and 0.7 mJ, respectively, were employed at 1 kHz pulse repetition rate. Their central wavelengths were $790$ nm for the FW input pulse and 1306 nm for the SW one, thus delivering FWM radiations at $\lambda_{\rm vis} = 566.3$ nm ($\nu_{\rm vis} = 529.8$ THz) and $\lambda_{\rm mid-IR} = 3764$ nm ($\nu_{\rm mid-IR} = 79.7$ THz). The expected secondary radiations assumed to emerge at frequencies $\nu_1\pm \nu_2$ (see Appendix \ref{sec:AppA} for an analytical justification) had thus to be centered around the wavelengths $\lambda_{\rm S1} = 2\,\mu$m ($\nu_{\rm S1} = 150$ THz), $\lambda_{\rm S2} = 1\,\mu$m ($\nu_{\rm S2} = 300$ THz) and $\lambda_{\rm S3} = 492$ nm ($\nu_{\rm S3} = 609.5$ THz). Special effort was carried out on the acquisition of angularly-resolved spectral distributions of such emissions selected by means of appropriate filters that blocked the main broadened pump and FWM components. Maximum available FW and SW energies were limited to 325 and 410 $\mu$J, respectively. Weak plasma fluorescence traces already occurred for FW energies only in excess of 0.03 mJ focused over 20 cm, so that adding a SW component with 0.02 mJ energy could contribute to the overall ionization at focus. Spectra were collected after the focused signal beam was transmitted through two scattering glass plates.\\

\begin{figure*}[ht]
  \centering \includegraphics[width=\textwidth]{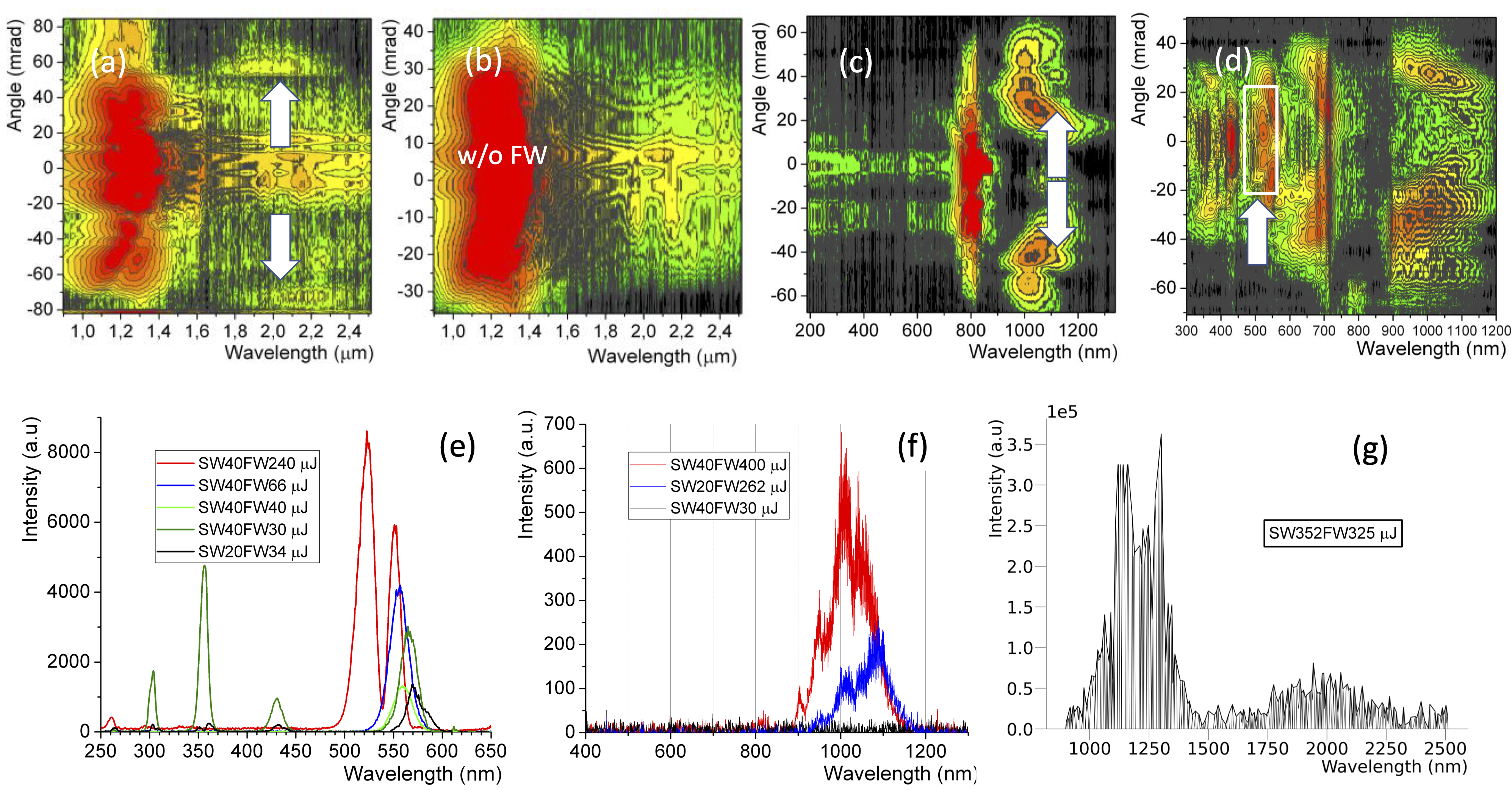}
  \caption{Secondary emissions measured from Vilnius's 20-cm-lensed setup occurring at the combined frequencies: (a,b) FW-SW ($\lambda_{S1}=2\,\mu$m) for (a) 0.32 mJ FW and 0.4 mJ SW energy but disappearing for (b) the 0.4 mJ SW beam alone, (c) 2FW-2SW ($\lambda_{S2}=1\,\mu$m) for 0.405 mJ FW and 0.23 mJ SW energy, and (d) FW+SW ($\lambda_{S3}=492$ nm - see white-edged rectangular area) for 0.1 mJ FW and 0.65 mJ SW energy. (e,f,g) Selected off-axis spectra (e) in the visible and (f) around 1 $\mu$m for various pump energies detailed in the legends. In (f) plasma-related signal near 495 nm is shown by red line (SW and FW pulse energies of 40 and 240 $\mu$J, respectively). (g) Off-axis spectrum registered near 2 $\mu$m (FW-SW) for 0.352 mJ SW and 0.325 mJ FW energy. Near 1.2 $\mu$m part of the off-axis SW spectrum is also plotted.}
\label{Fig5}
\end{figure*}

Figures \ref{Fig5}(a,b) demonstrate the existence of an expected secondary radiation located at $\lambda_{\rm S1} \simeq 2\,\mu$m wavelength and developing as out-off axis emission at +60 and -75 mrad when both SW and FW pulses are focused over 20 cm and form a plasma channel [compare Figs. \ref{Fig5}(a) and \ref{Fig5}(b)]. On-axis radiation in this spectral range is mainly produced by the broadened SW pulse alone, while the off-axis emission results from the common action of both SW and FW. When most of the visible radiations (pump and FWM components at about 570 nm) are blocked by filters, IR radiations can be registered in the range of $\lambda_{\rm S2} \simeq 1\mu$m, as detailed in Fig. \ref{Fig5}(c). This radiation was detected in the presence of air ionization. Figure \ref{Fig5}(d) shows an angularly-resolved frequency spectrum in the visible region for a SW pump being stronger in energy than the FW pulse (0.65 mJ against 0.1 mJ). We observe the secondary wavelength $\lambda_{\rm S3}$ close to the main FWM visible mode (see white rectangle). Similar angular-frequency spectral dynamics were collected when the FW pulse was more energetic than the SW one (0.405 mJ against 0.23 mJ, not shown), revealing this new component. Note the spectral signatures of the frequency combinations 2SW + FW and 3SW around their respective central wavelengths $357.6$ nm and $435.4$ nm.\\

Confirming these secondary radiations, Figs. \ref{Fig5}(e,f,g) furthermore display out-off axis spectra integrated over a small solid angle surrounding maximum emission zone for different FW and SW pulse energies. Besides contributions originating from high-order FWM-type combinations (3FW, 2FW+SW, 2SW+FW and 3SW), we observe the emergence of a peak close to FW+SW in Fig. \ref{Fig5}(e) ($S3$ radiation around 492-495 nm) connected to the visible mode at high enough energy in the FW and SW pulses (e.g., 240 and 40 $\mu$J, respectively). In Fig. \ref{Fig5}(f) the $S2$ signal around 1 $\mu$m is weak at this energy arrangement but it develops much more when the FW energy is increased to 400 $\mu$J. It has to be noticed that the two signals at $S3$ and $S2$ wavelengths disappeared in the absence or with too weak plasma traces. Figure \ref{Fig5}(g) details the off-axis spectrum registered near 2 $\mu$m radiation ($S1$ secondary radiation) for the SW and FW pulse energies of 0.352 and 0.325 mJ, respectively. A factor close to 10 distinguishes this secondary radiation from the main peak located at 1.2-1.3 $\mu$m and originating from the broadened SW pulse. Despite the latter peak, this secondary contribution is here clearly emerging.\\

\begin{figure*}[ht]
  \centering \includegraphics[width=0.8\textwidth]{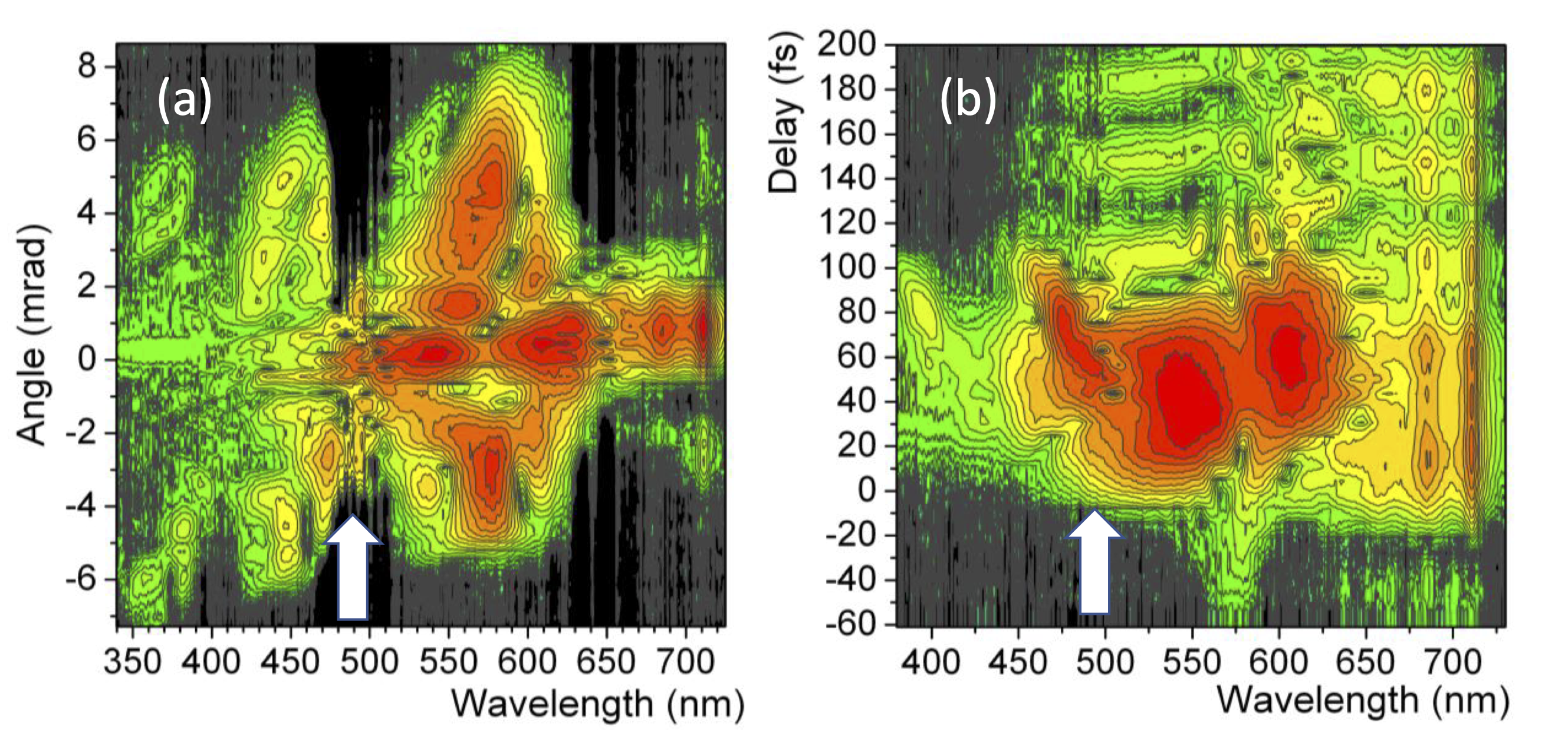}
  \caption{Secondary emission at FW+SW frequency ($\lambda_{S3} \simeq 492-495$ nm) near the generated visible radiation measured from Vilnius's 100-cm-lensed setup: (a) Frequency-angular spectrum of the generated visible emission near zero delay between the FW pulse with 0.41 mJ energy and the SW pulse with 0.545 mJ energy and (b) spectrum in the central bright spot of the emission as a function of the delay between the FW and SW pulses.}
\label{Fig6}
\end{figure*}

To validate the occurrence of the $S3$ (FW+SW) mode near the strong and broadened visible radiation, we also employed a filamentation setup with both pump components focused at $f = 1$ m. In this configuration, the $800$-nm FW and $1.3\,\mu$m SW beams with 35-45 fs duration propagated in loosely focused propagation starting from beam diameters $\approx$ 11 mm and 6 mm at 1/e$^2$ intensity level, respectively. Spectra were again collected after the collimated signal beam was transmitted through two scattering glass plates. Figure \ref{Fig6} reveals that out-off axis spectral components can develop independently of the 2SW + FW and 3SW frequency combinations at angles less than $\pm\,6$ mrad [compare with the broader angles $\pm\,20$ mrad reached in stronger focusing regime in Fig. \ref{Fig5}(d)] and preferably when the FW and SW pulses are slightly time delayed with each other, compared with the maximum signal acquired from zero delay by the FWM visible mode ($\lambda_{\rm vis} \simeq 578$ nm).

\section{Numerical simulations}
\label{sec3}

To clear up the generation process of both FWM and secondary radiations, we now perform numerical calculations, based, first, on the local current (LC) model \cite{Kim2008,Babushkin2011} to approach the radiated spectra produced by a two-color laser waveform subject to Kerr and/or plasma nonlinearities, second, on the 3D axis-symmetric unidirectional pulse propagation equation (UPPE) \cite{Kolesik2004} to moreover account for linear dispersion and nonlinear effects affecting the FW and SW pulses along filamentation.

\subsection{The local current (LC) model}
\label{sec3A}

We begin to examine the radiation spectra produced from the ELI laser parameters recalled in Table \ref{Table1}, using the simple local current (LC) model. The LC model is well-known to provide reliable trends of electromagnetic emissions created by air plasma nonlinearities, e.g., in the THz domain \cite{Kim2007,Kim2008,Nguyen2017,Nguyen2018NJP,Alirezaee2025}. It originally consists in first computing the free electron density, $N_e$, driven by field-induced ionization:
\begin{equation}
\label{Ne}
    \partial_t N_e = W(E) \left(N_a - N_e\right),
\end{equation}
where $N_a=2.7 \times 10^{19}$ cm$^{-3}$ is the initial neutral density and $W(E)$ is the field-dependent version of the photo-ionization rate derived by Perelomov, Popov and Terent'ev (PPT) \cite{Perelomov1966,Gonzalez2014}, adopting Talebpour {\it et al.}’s charge numbers $Z_{O_2}^* = 0.53$ for dioxygen (ionization potential $U_i = 12.1$ eV) and $Z_{N_2}^* = 0.9$ for di-nitrogen ($U_i = 15.6$ eV) \cite{Talebpour1999}. For air composed of 80$\%$ of N$_2$ and 20$\%$ of O$_2$, the electron density then follows from the differential equation:
\begin{equation}
\label{rateeqON}
\frac{\partial N_{O,N}}{\partial t}=W_{O,N}(E) (m_{O,N}N_a-N_{O,N}) ,
\end{equation}
where $W_{O,N}$ defines the ionization rate of either oxygen or nitrogen molecules with relative fractions $m_O=0.2$ and $m_{N}=0.8$, such that $N_e=N_O+N_N$. Photoionization can take place in the multiphoton ionization (MPI) regime at laser intensity levels limited to $\sim 10^{12-13}$~W/cm$^2$ for which $W(|E|) \approx \sigma_K |E|^{2K}$. Here, $\sigma_K$ is the MPI coefficient and $K=\mod{(U_i/\hbar \omega_1)}$ is the minimum number of photons needed to extract one electron from atmospheric species \cite{Keldysh1965,Perelomov1966}. Alternatively, at higher intensities $I> 10^{13}$ W/cm$^2$, photoionization takes place in the tunneling regime \cite{LandauQM1965,Ammosov1986,Berge2007} for which the laser electric field lowers enough the atom potential to let electrons tunnel through the Coulomb barrier. Free electrons are then accelerated by the laser field and generate photocurrents governed for non-relativistic plasmas by the current density $\vec{J}_e$:
\begin{equation}
\label{dtJe}
    \left(\partial_t + \nu_c \right) \vec{J}_e = \frac{e^2}{m_e} N_e \vec{E},
\end{equation}
where $e$, $m_e$ and $\nu_c$ denote the elementary charge, electron mass and an electron-neutral collision rate taken as $\nu_c = 2.85$~THz \cite{Tailliez2020}. Radiations generated by photocurrents usually refer to plasma-driven Brunel emissions \cite{Brunel1990}, whose spectra are formed by uneven integers of the fundamental frequency when only one pump component ionizes the medium or in various harmonic combinations when employing two colors with incommensurate frequency ratios \cite{Kostin2016,Wang2017,Zhang2017,David2025}. According to Jefimenko theory \cite{Jefimemko1989}, radiated electromagnetic emissions in the far-field mainly follow from the time derivative of the current density, i.e., $\vec{E}_{\rm rad} = g \, \partial_t {\vec J}_e$, $g$ denoting a geometrical factor $g=- \delta V/(4 \pi \epsilon_0 c^2 R)$ where $R$ is the (long) emitter-detector distance and $\delta V$ is the small emitter volume acting as a pointlike source.\\

For weaker laser intensities ($< 10^{12}$ W/cm$^2$), additional contributions to the overall radiated yield may be provided by the Kerr polarization induced from bound electrons, the instantaneous part of which reads as \cite{Borodin2013,Andreeva2016,Nguyen2017}:
\begin{equation}
\label{PKerr}
    \vec{P}_K(\vec{r}, t) = \varepsilon_0 \chi^{(3)} {E}^2(\vec{r}, t)\vec{E}(\vec{r}, t),
\end{equation}
with $\chi^{(3)}$ being the third-order electric susceptibility \cite{Boyd} assumed to be non-dispersive. This Kerr term is responsible for the transversal compression of the laser beam and maintains a filamentation dynamics through optical self-focusing and plasma defocusing sequences as its peak power remains higher than the critical power for self-focusing $P_K^{\rm cr} \approx 3.72 \lambda^2/(8 \pi n_0 n_2)$ \cite{Marburger,Berge2007}. Here, $n_0 = n_{\rm opt}(\omega)$ is the linear optical refractive index of air at frequency $\omega = 2 \pi c/\lambda$ taken from, e.g., Refs.~\cite{Peck1972,Ciddor1996} and $n_2=3\chi^{(3)}/(4n_0^2c\varepsilon_0)$ is the nonlinear refractive index. In addition, during filamentation in air, an intense laser pulse can also excite rotational transitions, mostly in $N_2$ molecules, leading to stimulated Raman rotational scattering (SRRS) with polarization vector:
\begin{equation}
\label{PRaman}
\vec{P}_R(t) = \frac{1}{2} \varepsilon_0 \chi^{(3)} \int_0^{+\infty} \vec{G}(\vec{E}, t -\tau) R(\tau) d\tau, \end{equation}
where the nonlinear vector $\vec{G}(\vec{E}, t, \tau)$ describes the third-order interaction and reduces to $\vec{G}(\vec{E}, t -\tau) = 3 E^2(t-\tau)E(t) \vec{e}_x$ for $x$-linearly polarized pulses (see Ref. \cite{Stathopulos2023}). $R(t) = \sin{(t/\tau_a)} \mbox{e}^{-t/\tau_b} (\tau_a^2 + \tau_b^2)/(\tau_a\tau_b^2)$ is the delayed Raman response function where $\tau_a \approx$ 62.5 fs and $\tau_b \approx$ 77 fs represent the rotational Raman and dipole dephasing times for $N_2$ molecules \cite{Penano2003,Pitts2004,Stathopulos2023}. This scattering nonlinearity affects the laser pulse propagation by delaying the nonlinear focus and smoothing the overall Kerr response. Its integral averages out the fast oscillations of the laser field  during a period of slow variation of the response function $R$. Thus, the Raman polarization is expected to mainly oscillate at pump frequency, and hence to barely contribute to the generation of additional frequencies, in particular to the THz part of the spectrum \cite{Nguyen2017,Stathopulos2023}. Accounting for the contribution of these Kerr nonlinearities in the far-field radiations leads, when assuming comparable volumes of the emitting sources, to extend the Jefimemko model for the radiated field to optical nonlinearities as
\begin{equation}
    \label{dtJtotal}
    \vec{E}_{\rm rad} = g \left(\partial_t J_e + \partial_t^2 [ (1-x_K) \vec{P}_K + x_k \vec{P}_R ] \right).
\end{equation}
where $x_K$ is the ratio of SRRS to the total third-order response, estimated to $\approx 80$~\% according to~\cite{Zahedpour2015}.\\

In the following we shall consider linearly polarized, parallel pump components, directed along the $\vec{x}$ axis and having Gaussian temporal profiles such as, e.g., 
\begin{equation}
\label{ELPP}    
\vec{E}_L(t) = \left[E_1 \mbox{e}^{- 2 \ln{2} \frac{t^2}{\tau_1^2}} \cos(\omega_1 t) + E_2 \mbox{e}^{- 2 \ln{2} \frac{t^2}{\tau_2^2}} \cos(\omega_2 t + \phi)\right] \vec{e}_x, 
\end{equation}
$\phi$ denoting the relative phase between the FW and SW pulses centered on the frequencies $\omega_1 = 2 \pi \times 374$ THz and $\omega_2 = 2 \pi \times 232$ THz, respectively. $\tau_1 = 35$ fs and $\tau_2 = 40$ fs refer to the FW and SW pulse durations (FWHM) employed during the ELI experiment and $E_1,E_2$ are the corresponding input electric fields (in V/m) determined by the FW and SW input energies. Plugging the two-color field Eq. (\ref{ELPP}) into Eq. (\ref{rateeqON}) and computing the electron current Eq. (\ref{dtJe}) and bound electron nonlinearities [Eqs. (\ref{PKerr}) and (\ref{PRaman})] allows us to evaluate the radiated field Eq. (\ref{dtJtotal}). A Fourier transform in time then yields the corresponding radiation spectrum.\\

\begin{figure*}[ht]
  \centering \includegraphics[width=\textwidth]{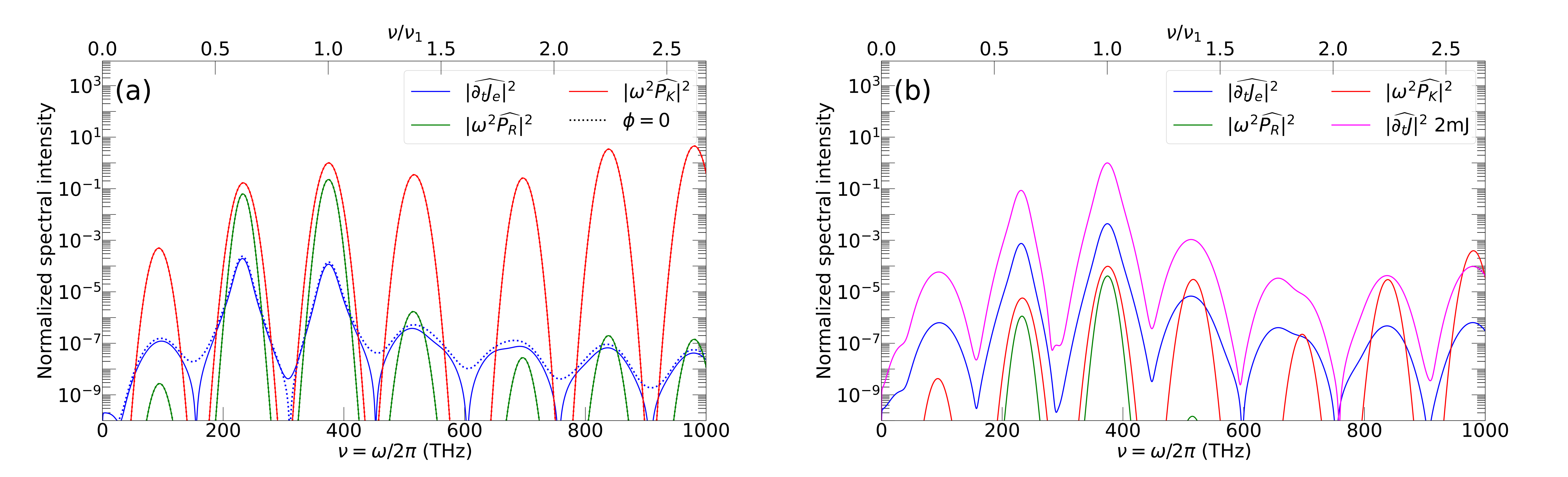}
 \centering \includegraphics[width=\textwidth]{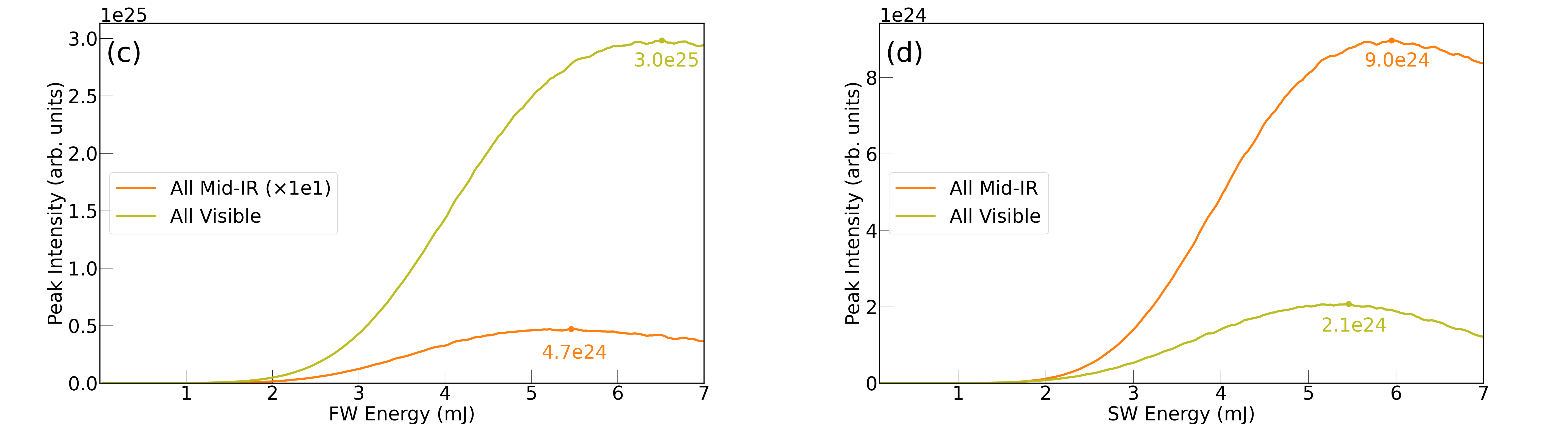}
  \centering \includegraphics[width=\textwidth]{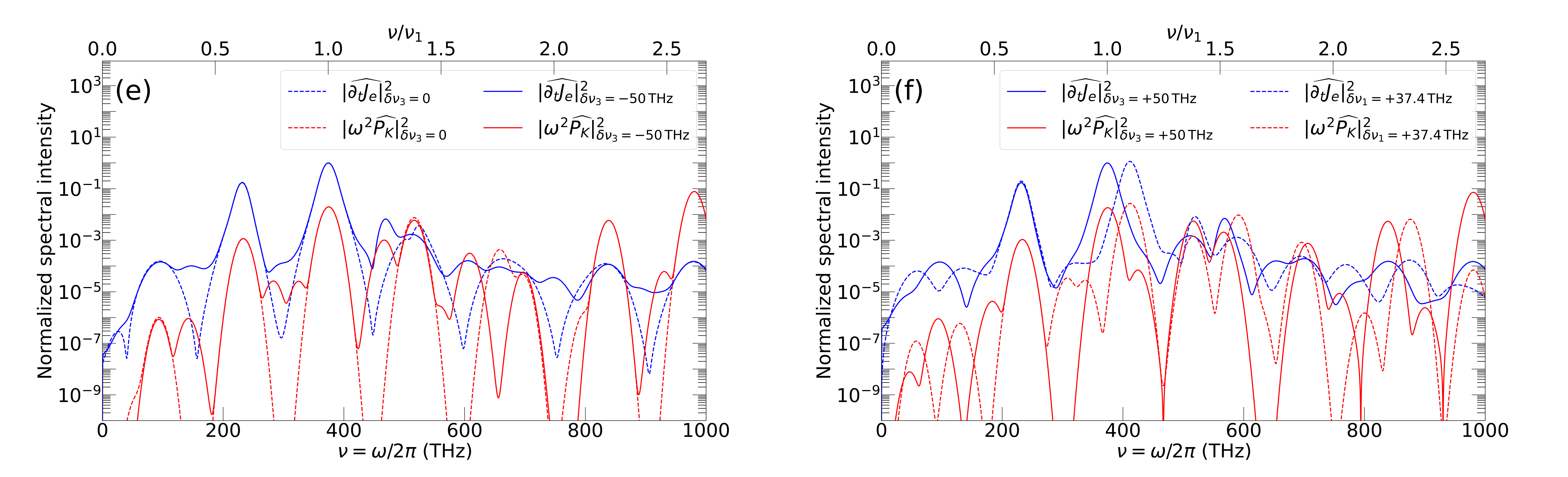}
  \caption{(a) Radiated spectra computed from the LC model with instantaneous Kerr (red curve), Raman-delayed (green curve), and PPT-driven plasma response (blue curve) for a relative phase $\phi= \pi/2$ (solid curves) created by a 0.1-mJ FW pulse and a 0.13 mJ SW pulse. Dotted curves show the same data for a zero relative phase. (b) Same results for a 1-mJ FW pulse and for a 2-mJ FW pulse using the PPT rate only (purple curve). (c,d) Radiated peak intensity (in arb. units) of the visible and mid-IR modes function of (c) the FW pulse energy and (d) the SW pulse energy computed from all Kerr and plasma sources over the energy range $<7$ mJ. (e,f) Influence of a third visible component for 1 mJ FW energy and 0.13 mJ SW energy (e) without or with a negative detuning $\delta \nu_3$ (in THz) and (f) with a positive detuning for $\theta=0$ in the visible pulse phase [see Eq. (\ref{E2})]. In (f) the dashed curve shows spectral distortions induced by positive detunings both in the visible frequency $\delta \nu_3 = + 50$ THz and in the FW frequency $\delta \nu_1/\nu_1 = + 37.4$ THz. In (b-f), the relative phase $\phi$ between the FW and SW pulse is equal to $\pi/2$.}
\label{Fig7}
\end{figure*}

Two examples of such radiation spectra are shown in Figs. \ref{Fig7}(a) and \ref{Fig7}(b) for the input FW and SW fields with $E_1 = 27.45/\sqrt{10}$ GV/m [Fig. \ref{Fig7}(a)] or $E_1 = 27.45$ GV/m [Fig. \ref{Fig7}(b)] and $E_2 = 9.88$ GV/m, for the relative phase $\phi = \pi/2$. These field values are associated with FW and SW energies of 0.1 mJ or 1 mJ and 0.13 mJ, respectively. With an FW intensity close to 100 TW/cm$^2$ [1 mJ FW energy, see Fig. \ref{Fig7}(b)], ionization takes place mainly in the tunnel regime through the fundamental pump pulse only. From this couple of figures one can observe that the FWM radiation modes are conveyed mainly by the instantaneous Kerr nonlinearity (compare red and green curves) at low FW energy and the contribution of SRRS to the overall radiation yield appears negligible as the SRRS integral in time averages out new radiation besides the pump components. Changing the value of the relative phase $\phi$ from $\pi/2$ to 0 [Fig. \ref{Fig7}(a)] does not really affect the spectral distributions as already reported in \cite{Kim2009}. When increasing the FW pulse intensity above the tunnel ionization threshold [Fig. \ref{Fig7}(b)] one clearly sees that ionization (blue curve) may take over the Kerr-driven FWM generation process, at least in the mid-IR range (compare blue and red curves). Increasing the FW energy to 2 mJ renders ionization dominant in the production of visible radiation [see purple curve in Fig. \ref{Fig7}(b)]. Compared with the Kerr-driven spectrum, PPT ionization creates a pedestal induced by the attosecond-scaled ionization events and associated stepwise increase in the electron density, which produces broader bandwidths \cite{Babushkin2011,David2025}. \\

Figures \ref{Fig7}(c) and \ref{Fig7}(d) plot the growth in the mid-IR and visible intensity (in arb. units) inferred from the peaks of the FWM spectral modes as a function of the FW [Fig. \ref{Fig7}(c)] and SW [Fig. \ref{Fig7}(d)] pulse energy, computed from the overall nonlinearity including both Kerr and plasma responses over an extended energy range. One sees that these gain curves first increase, reach a maximum and then decrease from pump energies $>5$ mJ, which we attribute to the saturation in the photoionization efficiency and destructive interferences between Kerr and plasma contributions.\\

It is then instructive to compare such energy growths to their experimental counterparts, as presented in Figs. \ref{Fig3} and \ref{Fig4}. Here, instead of plotting ''best'' curves fitting our experimental data, we fix every emitting source (instantaneous Kerr response, plasma response alone and sum of all nonlinearities) to an experimental reference point defined by the 79 nJ yield measured at 0.11 mJ SW energy and 1 mJ FW energy. The energy yield at 0.1 mJ has been decreased by a factor 10 accordingly. From these figures we can observe that at low FW energy $\sim 0.1$ mJ [see Figs. \ref{Fig3}(a) and \ref{Fig4}(a)], both mid-IR and visible emissions mainly proceed from the instantaneous Kerr contribution, as expected, with a logical quadratic increase for the former radiation and a linear one for the latter as function of the SW energy. When the FW energy is increased to 1 mJ, the visible emission goes on being driven by the Kerr response $\sim E_{FW}^2 E_{SW}$, which remains compatible with Fig. \ref{Fig7}(b) and exhibits linear (quadratic) energy increase with the SW (resp. FW) pump energy. Surprisingly, the mid-IR radiation follows the same trend, i.e., unlike the LC expectations, the Kerr response appears still dominating among the possible emitters and also follows Kerr-type behavior $\sim E_{SW}^2 E_{FW}$ with net quadratic (linear) energy increase when varying the SW (resp. FW) pulse energy. Thus, in the pump energy range investigated experimentally, a Kerr-scaling applies up to FW energies limited to about 1-2 mJ. Beyond this limit, the plasma response may take over (see green solid curves in Figs. \ref{Fig3} and \ref{Fig4}). However, our experimental measurements [Figs. \ref{Fig3}(c) and \ref{Fig4}(c)] display a net saturation in the FWM energy growth, which rather follows that illustrated around 5-6 mJ in Figs. \ref{Fig7}(c,d), although at weaker pump energies.\\

Furthermore, it is clear from Figs. \ref{Fig7}(a) and \ref{Fig7}(b) that no secondary radiation occurs at the expected frequency combinations (FW-SW), 2(FW-SW) and (FW+SW), which may need to account for the influence of other spectral components than the pump ones, emerging, e.g., through nonlinear propagation effects. Addressing this issue, Figs. \ref{Fig7}(e) and \ref{Fig7}(f) detail spectral distortions induced by the coupling of the two input colors with the visible radiation, which appears as the main player able to interact with the pump fields. FW and SW pulse energies are kept at 1 mJ and 0.13 mJ, respectively. Only spectra associated with the plasma and instantaneous Kerr response are displayed here. These spectra are consistently evaluated from the three-wave input field: 
\begin{equation}
\label{E2}    
E_L(t) = E_1  \cos(\omega_1 t) + E_2 \cos(\omega_2 t+\phi) + E_3 \cos((\omega_3 + \delta \omega_3)t + \theta), 
\end{equation}
with constant amplitudes $E_j,\,j=1,2,3$, relative phases $\phi$ and $\theta$, and a visible frequency $\omega_3 = 2\pi \nu_{3}$ detuned by $\delta \omega_3 = 2 \pi \delta \nu_{3}$. Interestingly, Fig. \ref{Fig7}(e) shows that the visible component strictly defined at frequency $\nu_{\rm vis} \equiv \nu_3 = 2 \nu_1-\nu_2 = 516$ THz (dashed blue curve) does not excite secondary frequencies and preserves the same spectral modes as in Fig. \ref{Fig7}(b). Let us now examine the effect of frequency shifts associated with reasonable spectral broadenings, i.e., $\delta \nu/\nu \leq 10\%$. When we insert a negative detuning and decrease the visible frequency such as $\nu_{\rm vis} \rightarrow \nu_{\rm vis} + \delta \nu_{\rm vis}$ with $\delta \nu_{\rm vis} \equiv \delta \nu_3 = -50$ THz, strong spectral distortions develop as the visible mode does no longer satisfy the resonance conditions responsible for the FWM modes. From the solid curves of Fig. \ref{Fig7}(e), we observe that -- i/ both Kerr and plasma nonlinearities broaden more their respective spectra than in the absence of detuning, particularly in the spectral bandwidths $100-200$ THz, around 300 THz, and between 450 THz and 650 THz, hosting the secondary radiations $\nu_1\pm\nu_2$ and $2(\nu_1-\nu_2)$ -- ii/ with 1 mJ FW energy, spectral broadening appears driven in the mid-IR region by the plasma contribution although the latter may interplay with Kerr-induced fluctuations in the visible range. Figure \ref{Fig7}(f) illustrates the role of positive detunings $\delta \nu_{3} = + 50$ THz (solid curves), which cause the excitation of new frequencies around 600 THz and the emergence of THz components in the limit $\nu \rightarrow 0$, while the Kerr spectrum seems more preserved (solid curves). Interestingly, introducing $10\%$ detuning in the FW wavelength ($\delta \nu_{3} = + 37.4$ THz), standing for pump broadening effects, also contributes to the development of the searched secondary radiations (see dashed curves). Changing the relative phase $\theta$ of the visible mode from 0 to $\pi/2$ introduces minor modulations that could enhance the spectral amplitude of the secondary radiation (not shown). As detailed in Appendix \ref{sec:AppA}, analytical calculations based on a three-wave model using a multiphoton ionization rate, $W(t) \propto |E(t)|^{2K},\,K \geq 1$, allows us to justify that at fixed FW and SW frequencies, the generated visible component must be detuned to some extent to trigger secondary radiations, may a plasma regime ($K \gg 1$) or a Kerr one ($K=1$) be considered.\\

In summary, LC computations indicate that plasma generation broadens FWM radiations and may also increase their energy yields, besides the action of optical Kerr nonlinearities. However, direct confrontations of LC results with our experimental data show that the Kerr stage and related energy scalings should drive the generation process for the main FWM modes when the dominant FW pump pulse contains energies limited to $\sim 1-2$ mJ. Beyond this threshold, saturation effects prevent the conversion process, which the LC model predicts in fact at higher pump energies [$\sim 5$ mJ, see Figs. \ref{Fig7}(c,d)]. Importantly, secondary radiations and frequency combinations (FW $\pm$ SW) and their multiples can be justified from the LC model and should develop only if the FWM visible mode is broad enough.

\subsection{The Unidirectional pulse propagation equation (UPPE) model}
\label{sec3B}

To investigate the dynamics of the FWM and secondary radiations along the laser beam propagation, including transverse and nonlinear propagation effects, we now perform three-dimensional comprehensive simulations based on the UPPE \cite{Kolesik2004,Vaicaitis2018} for both the ELI filamentation setup and the Vilnius setup in focused geometry. The UPPE model for $x-$linearly polarized two-color pulses governs the forward rapidly-oscillating component of the optical electric field as described by
\begin{equation} 
\label{UPPE}
    \frac{\partial \overline{E}}{\partial z} = i \left(\sqrt{k^2(\omega) - k_\perp^2} - \frac{\omega}{v_g}\right) \overline{E} + \frac{i \mu_0 \omega^2}{2 k(\omega)}  \left[\frac{i}{\omega} \left(\overline{J}_e + \overline{J}_{\rm loss} \right) + \overline{P}_{K} \right].
\end{equation}
As we assume cylindrical symmetry, bar notation refers to both Fourier transform in time and Hankel transform in the radial coordinate $r = \sqrt{x^2+y^2}$ of any function $f(r,z,t)$. $k(\omega) = n(\omega) \omega/c$ is the pulse frequency depending on the linear refractive index of air based on Ref. \cite{Ciddor1996} and $v_g$ is the FW group velocity \cite{Couairon2011}. $\overline{J}_{\rm loss}$ denotes the loss current density associated with photoionization \cite{Berge2007} that drives the current density [Eq. (\ref{dtJe})]. For pulse durations as short as 35 fs avalanche ionization and electron recombination are neglected. Photoionization is described using the same PPT rate as in Sec. \ref{sec3A}. The two-color input field is chosen with Gaussian profiles both in space and time. At $z=0$, the optical field is described by the sum of the FW (index $j=1$) and SW (resp. $j=2$) field components:
\begin{equation}
\label{input}
E(t,r,z=0)= \sum_{j=1,2} E_{j,0} \exp\left(-\frac{r^2}{r_{j0}^2}\right) {\cal F}^{-1} \left[{\cal F} \left(\exp\left(-2\ln{2}\,\frac{t^2}{\tau^2} -i\omega_{j}t\right) -i\frac{\omega r^2}{2f_jc}\right)\right],
\end{equation}
where symbol $\cal{F}$ means Fourier transform in time, $\tau=35$ fs is the FWHM pulse duration which, for technical convenience, we take equal for the two colors having 1/e$^2$ input beam radius $r_{j0}$, focal length $f_j$, input amplitude $E_{j,0}$ and central angular frequency $\omega_{j}$. Because the Raman-delayed Kerr contribution does not generate FWM radiations, we only consider an overall instantaneous Kerr response with a mean nonlinear refractive index $n_2 = 10^{-19}$ cm$^2$/W, corresponding to a critical self-focusing power of $P_{\rm cr} \approx 9.5$ GW for a $\sim\,800$-nm FW pulse. With FW energies limited to 1 mJ and thus containing about 3 critical powers, we do not expect any multiple filamentation dynamics, even in loosely focused propagation \cite{Berge2007}, which thereby justifies the use of an axis-symmetric UPPE model.\\

Our UPPE numerical model has been employed to simulate both the ELI filamentation setup introduced in Sec. \ref{sec2} (see laser parameters, associated pump and generated frequencies in Table \ref{Table1}, top) and the Vilnius setup in focused propagation (resp. Table \ref{Table1}, bottom). Equation (\ref{UPPE}) was integrated using the Fourier split-step method \cite{Wartak2013} and, due to azimuthal symmetry, fast Hankel transforms \cite{Hankel2005}. For the ELI setup, the FW pulse was first beforehand propagated linearly from $z=0$ to $z=21$ cm; the SW and FW pulse components were then both propagated from the latter distance to their common focal distance $z = f_1 = 96$ cm. Their relative phase at $z=21$ cm was generically set as $\phi =0$, since, in accordance with \cite{Kim2009} and Fig. \ref{Fig7}(a), we preliminarily verified the absence of significant impact of this quantity onto the FWM emission process. Starting with $\Delta z = 1\,\mu$m, the longitudinal increment was self-adaptive, reducing with the maximum intensity and peak plasma density near focus. Maximum resolution was 0.18 fs in time, $5.45\,\mu$m in transverse coordinate, 0.2 $\mu$m in $z$ and 0.67 THz in spectral step.\\

\begin{figure*}[ht]
  \centering \includegraphics[width=0.9\textwidth]{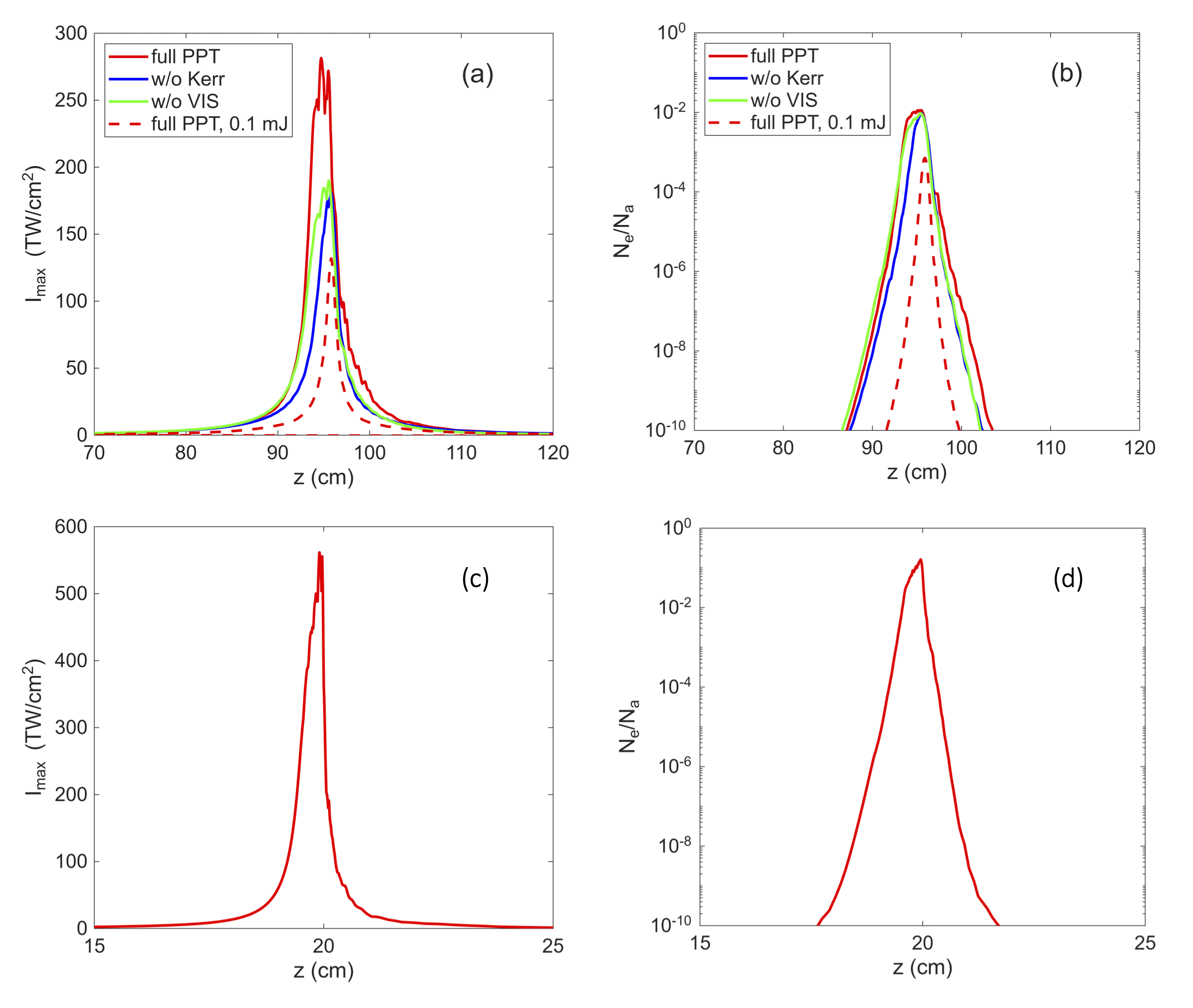}
  \caption{(a,b) Pulse propagation aspects for the ELI setup: (a) Maximum intensity of a two-color filament with a 1 mJ FW pulse coupled with a 0.13 mJ SW pulse produced by the ELI setup. (b) Corresponding peak plasma density $N_e/N_a$ normalized to the neutral atom density with (red curve) or without the Kerr response (blue curve) and without the visible frequency region $\nu > 470$ THz (green curve) for a relative phase $\phi = 0$ at the entrance of the SW beam $(z = 21$ cm). The red dashed curves show the same quantities for a 0.1 mJ FW pulse. (c,d) Same information simulating all nonlinearities for the Vilnius setup at focal distance $f=20$ cm.}
\label{Fig8}
\end{figure*}

UPPE results summarizing pulse propagation aspects are displayed in Figs. \ref{Fig8}(a,b) for the ELI setup, which shows the maximum intensity [Fig. \ref{Fig8}(a)] and peak plasma density [Fig. \ref{Fig8}(b)] evolving along more than one meter length for Gaussian pump components with 0.13 mJ SW energy and 1 mJ (solid curves) or 0.1 mJ (red dashed curve) FW energy. For 1 mJ FW energy, these quantities are plotted when accounting (red curves) or not (blue curves) for the Kerr response and when one filters out at each $z$-step the generated visible radiation ($\nu > 470$ THz) from the nonlinearities (green curves). At the distance $z = 50$ cm it is clear that the combination of the two colors does not ionize air as the pulse intensity remains at lower values than 1 TW/cm$^2$, thus privileging Kerr-driven FWM radiations as expected from Fig. \ref{Fig7}(a). In contrast, a plasma channel with mean density $\sim 10^{17}$ cm$^{-3}$ emerges around $z = f_1$ for the 1 mJ FW pulse, and it develops upon $\sim 5$ cm in air, in agreement with the plasma fluorescence traces displayed in Fig. \ref{Fig1}(b). When the Kerr term is removed from the overall propagation dynamics (blue solid curves), the beam focus is slightly shifted forward along the propagation axis. By comparison with the red curve, including Kerr self-focusing locally amplifies the plasma yield, which, by feedback, increases the maximum clamping intensity to some extent. By comparison, removing all frequencies from the visible bandwidth $\nu > 470$ THz using a low-pass frequency filter diminishes the energy content of the beam, which then reaches a weaker intensity at focus. In the case of a 0.1 mJ FW pulse energy (red dashed curve), we observe that the beam propagation is close to linear, i.e. maximum intensity is exactly reached at linear focus, while the peak electron density is one to two orders of magnitude below that produced by the 1 mJ FW pulse and thus barely exceeds the common detection threshold $\sim 10^{15}$ cm$^{-3}$. Figures \ref{Fig8}(c,d) show similar behaviors for the Vilnius setup along shorter propagation distances in focused geometry: maximum pulse intensity [Fig. \ref{Fig8}(c)] is attained at focus ($z \simeq f = 20$ cm), where a short ($\lesssim 1$-cm long) plasma filament develops [Fig. \ref{Fig8}(d)].\\

\begin{figure*}[ht]
  \centering \includegraphics[width=0.9\textwidth]{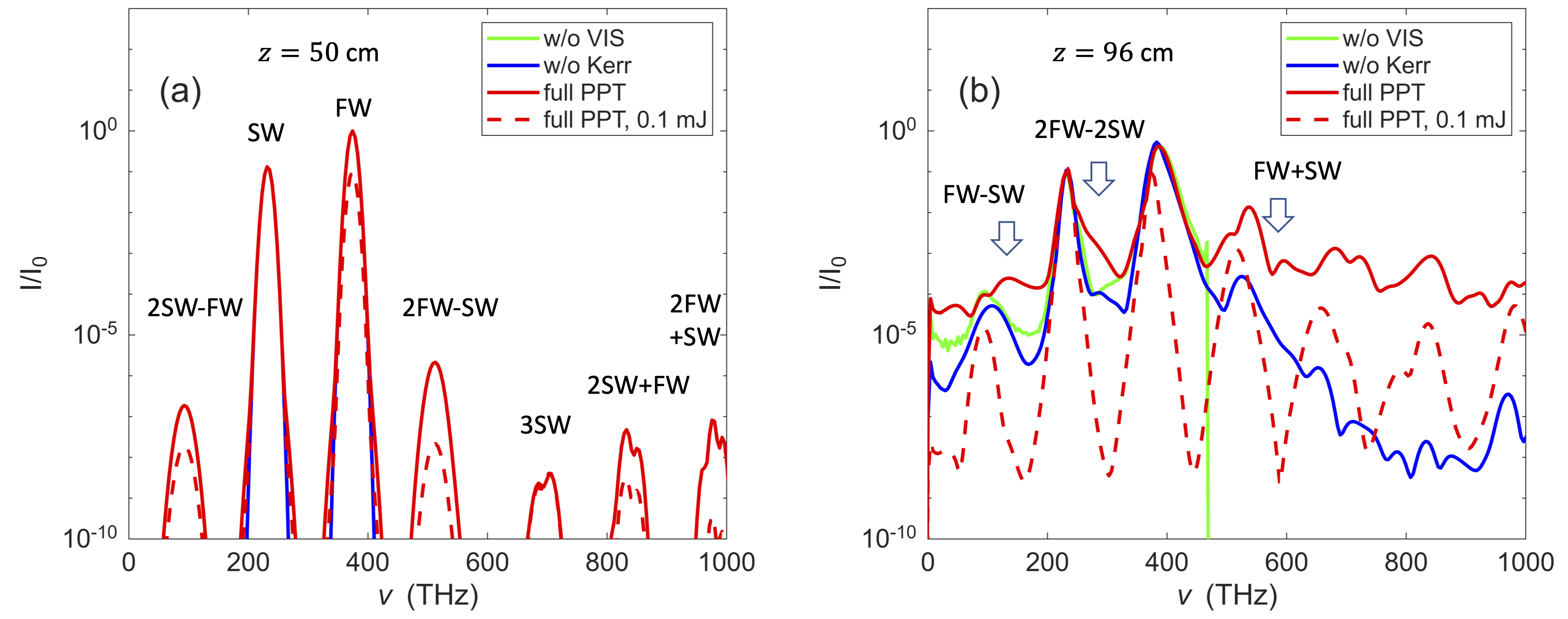}
   \centering \includegraphics[width=0.9\textwidth]{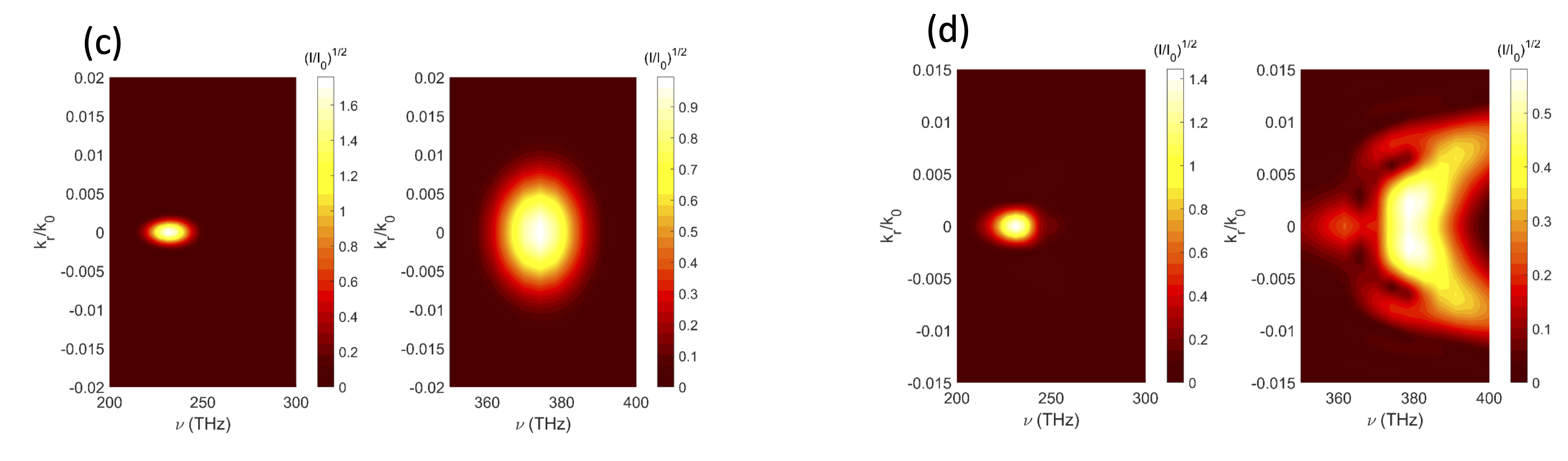}
  \caption{(a,b) Angle-integrated spectra computed from UPPE and normalized to the input FW spectrum at (a) $z = 50$ cm where there is no significant plasma generation and (b) $z =96$ cm with significant plasma generation by the 1-mJ FW pulse (red solid curves). Red dashed curves refer to the 0.1 mJ energy FW pulse (dashed curves). Pump and generated frequencies are indicated. (c,d) Angle-resolved spectra normalized to the input FW spectrum zooming in on the pump frequency regions at (c) $z = 50$ cm and (d) $z = 96$ cm.}
\label{Fig9}
\end{figure*}

In addition, integrated spectra normalized to the input FW spectrum are presented for the ELI setup in Fig. \ref{Fig9} at $z=50$ cm [Fig. \ref{Fig9}(a)] and $z = f_1 = 96$ cm [Fig. \ref{Fig9}(b)] for the two cases of 0.1 mJ and 1 mJ FW energy. These spectra were calculated as $I(\nu)=\int_0^{\infty}|S(\nu,k_r)|^2k_r dk_r$. $S(\nu,k_r)$ being the spectral amplitude and $k_r=\sqrt{k_x^2+k_y^2}$. At $z=50$ cm (low intensity regime) only the FWM modes expected from the Kerr response develop at the frequency combinations 2SW-FW, 2FW-SW, 3SW, 2SW+FW and 2FW+SW. They, of course, achieve much weaker intensity with the 0.1 mJ FW input pulse. By contrast, around the FW focus (high intensity regime) for the 1 mJ pump pulse, not only similar modes superimposed to the Kerr-driven ones, but also a typical pedestal signaling ionization events (see previous subsection) occur, together with the emergence of weaker, secondary radiations located near the frequency combinations FW-SW, 2FW-2SW and FW+SW (see Table \ref{Table1}). Importantly, repeating the same simulation while ignoring the Kerr response demonstrates that an efficient Kerr stage is necessary both to seed and amplify the FWM modes before the plasma response comes in and to trigger the secondary radiations in the plasma regime. The green curve shows the same UPPE data after filtering out the broadened visible bandwidth and higher frequencies from the plasma and Kerr nonlinearities all along the beam propagation. We can clearly see that suppressing the overall generated visible components prevents from creating the THz and secondary radiations developed in the red curve. In other words, the FWM-triggered visible mode is crucial in the generation process of the weaker secondary radiations, as argued above. The overall frequency bandwidth around the generated visible wavelength (2FW-SW) attains 25 THz in the Kerr regime ($z=50$ cm) and up to 37 THz in the plasma regime ($z=95$ cm) (not shown), within the below-$10\%$ bandwidth chosen in our LC calculations (see subsection \ref{sec3A}).\\

Furthermore, Figs. \ref{Fig9}(c,d) show the angle-resolved spectra $k_r/k_0$ in their respective frequency intervals. Such frequency spectra are here displayed in amplitude for accessing more detailed features in the secondary emissions addressed below. All spectra are normalized to the input spectral intensity $I_0 = 1.5 \times 10^{-20}$ J.s.m$^2$, corresponding to the incident FW intensity of $\sim 2.65\times 10^{14}$ W/cm$^2$ (for 1 mJ FW energy). Note the narrower spatial spectral extent of the SW component: as both FW and SW beams have spatial chirp due to lens focusing, their spatial Fourier transform behaves in intensity like $\mbox{e}^{-k_r^2 (\lambda_j f_j)^2/(2 \pi^2 r_{j0}^2)}$ [using notations of Eq. (\ref{input}) together with $\lambda_j f_j \ll r_{j0}^2$], reducing thereby the angular spectrum of the SW wave $(j=2)$ with smallest beam radius. An important property revealed by these patterns is that the SW intensity takes over the FW intensity near the focus $z = f_1$, which we attribute to mutual nonlinear couplings between the two pulse components through, e.g., cross-phase modulation.\\

\begin{figure*}[ht]
 \centering \includegraphics[width=0.9\textwidth]{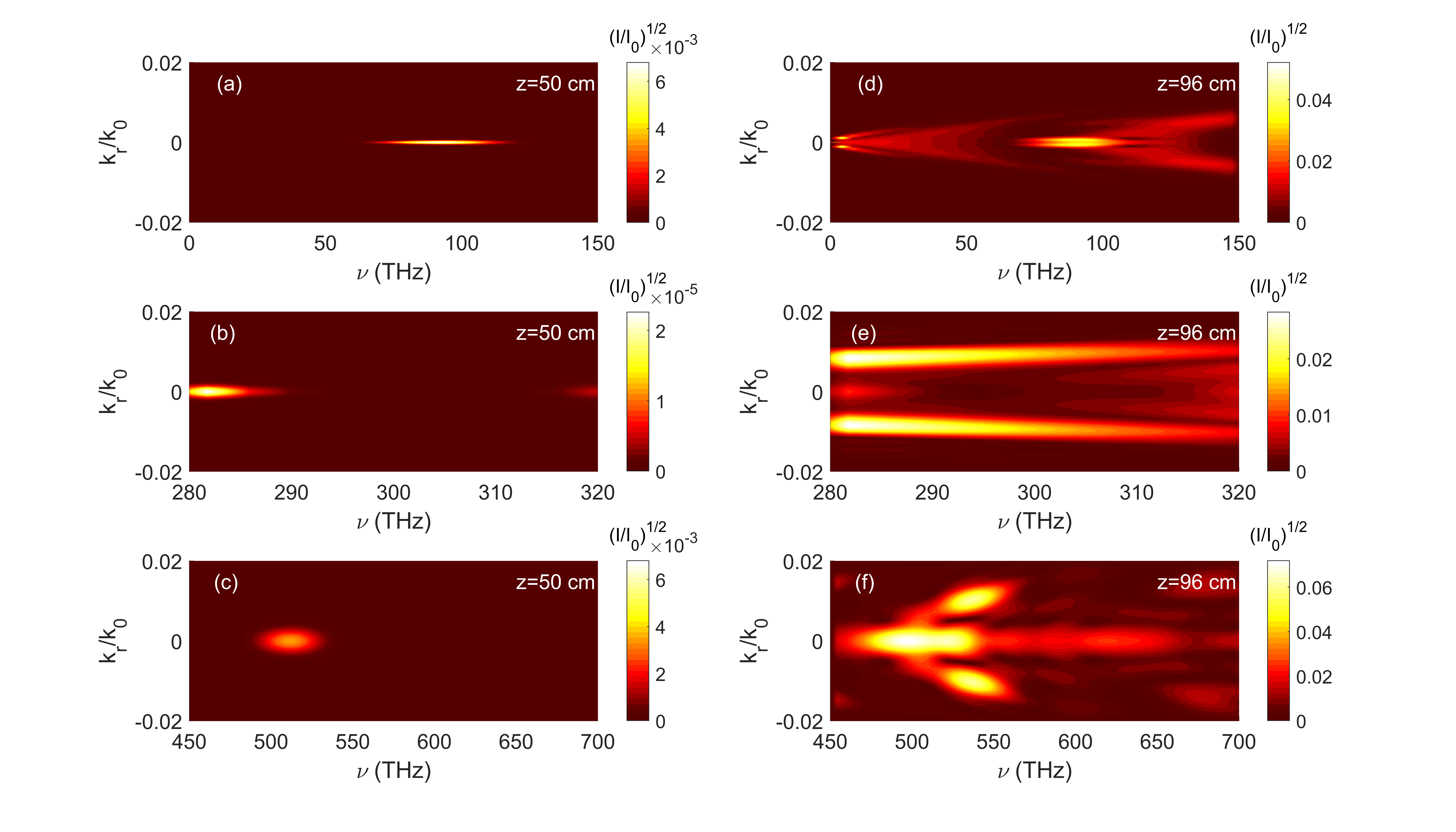}
  \caption{Angle-frequency spectra normalized to the maximum FW intensity at $z=0$ computed with UPPE and detailing secondary emissions in the THz-mid-IR bandwidth, around 300-THz and in the visible range at (a-c) $z = 50$ cm where there is no significant plasma generation and (d-f) $z = 96$ cm with significant plasma generation induced with 1-mJ FW energy and zero relative phase between the input FW and SW pulses at the entrance of the SW beam $z = 21$ cm.}
\label{Fig10}
\end{figure*}

\begin{figure*}[ht]
 \centering \includegraphics[width=0.9\textwidth]{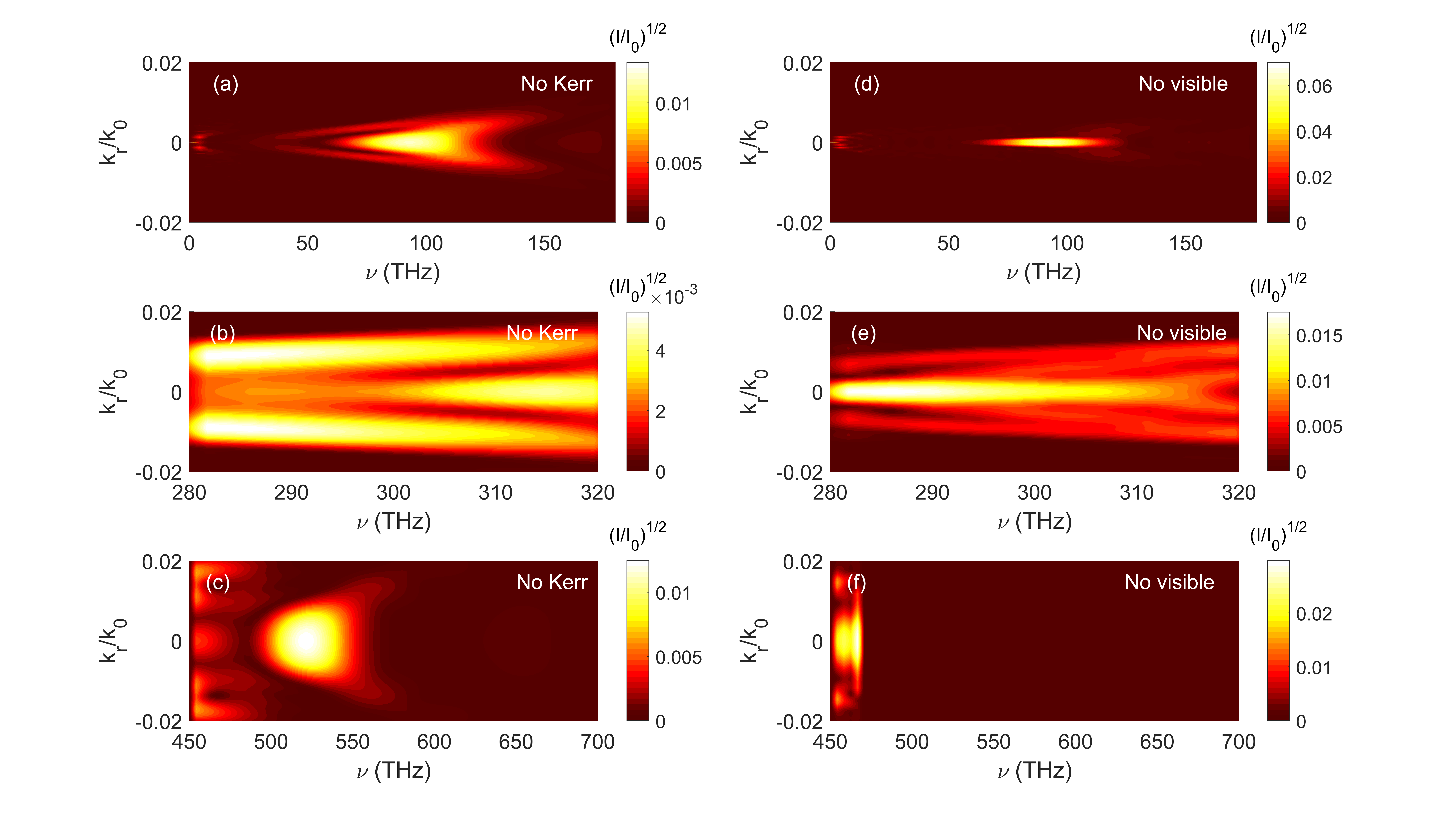}
  \caption{Same angle-frequency spectra as in Fig. \ref{Fig10} at $z = 96$ cm but (a-c) without the Kerr response and (d-f) without the broadened visible region [see cut-off in Fig. \ref{Fig9}(b)].}
\label{Fig11}
\end{figure*}

Next, we show the frequency-angular spectra from selected emission bandwidths in Fig. \ref{Fig10}. In the Kerr stage [$z=50$ cm, Figs. \ref{Fig10}(a-c)], no secondary radiation occurs. By contrast, in the plasma stage [Fig. \ref{Fig10}(d-e)], we observe ring formation in the mid-IR spectrum followed by secondary radiation emitted off-axis around 142 THz, which corresponds to the frequency combination FW-SW. Also, the bandwidth 280-320 THz becomes active which we attribute to the cascading process 2(FW-SW), although partly overlapping with the broadened SW pulse. Note that higher-order frequencies near 4(FW-SW) contribute to the off-axis lobes in Fig. \ref{Fig10}(f). The expected signal at around 606 THz associated with the combination FW+SW clearly develops, mainly on-axis. Terahertz frequencies (close to zero) result from photocurrents induced by the ionizing (dominant) broad SW pump pulse $\sim 225-240$ THz [see Fig. \ref{Fig9}(d)] coupled with the non-ionizing broadened visible radiation $\leq 550$ THz close to the FW focus. For completeness, Fig. \ref{Fig11} displays the same frequency-angular spectra produced without the Kerr response or without the broadened visible radiation. A direct comparison with Figs. \ref{Fig10}(d-f) confirms that -- i/ the Kerr nonlinearity conditions the spectral dynamics at focus, in particular that of most expected secondary radiations and -- ii/ the latter cannot develop in the absence of the visible mode.\\

\begin{figure*}[ht]
  \centering \includegraphics[width=0.9\textwidth]{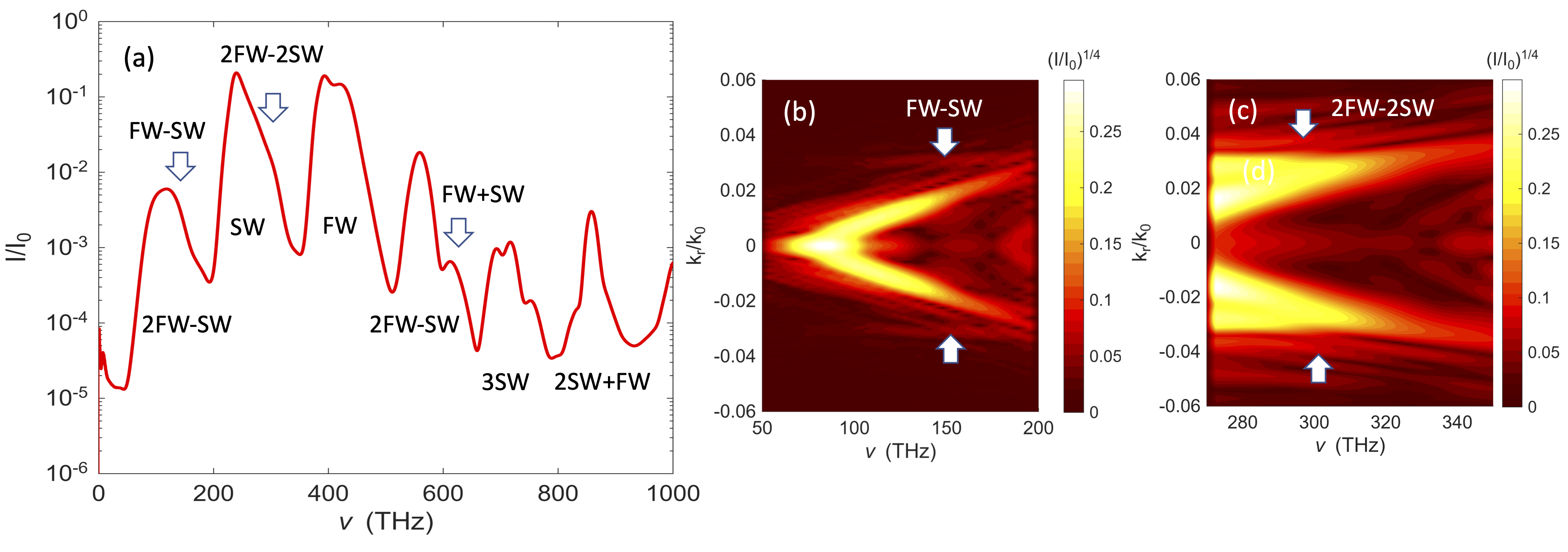}
  \centering \includegraphics[width=0.9\textwidth]{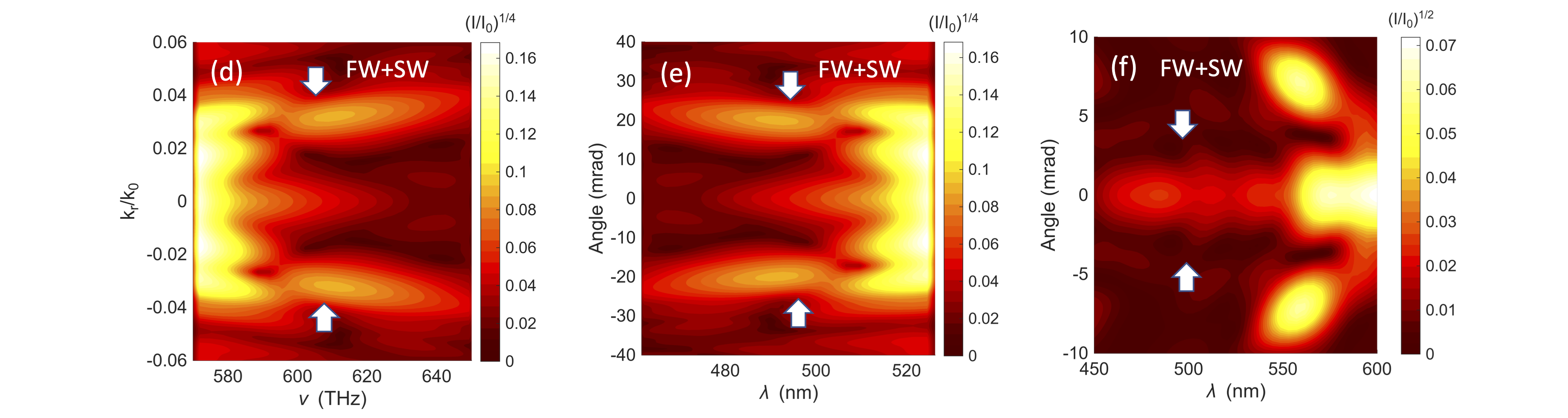}
  \caption{(a) Angle-integrated spectrum computed with UPPE and normalized to the input FW spectrum at $z = 20$ cm for the Vilnius setup. (b-d) Angle-resolved spectra normalized to the input FW spectrum zooming in on the secondary radiation frequency of interest, i.e., (b) FW-SW, (c) 2(FW-SW) and (d) FW+SW at the same propagation distance. (e,f) Angular spectrum computed from $[k_r \nu_0/(k_0 \nu)]$ of the FW+SW radiation function of the wavelength for (e) the Vilnius setup and (f) the ELI setup.}
\label{Fig12}
\end{figure*}

To end with, Fig. \ref{Fig12} details similar features in the angle-integrated spectra and the angle-resolved spectra extracted from the UPPE simulation of the Vilnius setup. Due to the tighter focusing (20-cm focal length) of the two colors, plasma-induced broadening appears less pronounced [compare Figs. \ref{Fig12}(a) and \ref{Fig9}(b)]. The integrated spectrum in Fig. \ref{Fig12}(a) confirms the occurrence of weaker, secondary radiations located at the frequency combinations 2(FW-SW) and FW+SW. Note, however, that, compared with Fig. \ref{Fig9}(b), the expected FW-SW mode seems to merge with the mid-IR radiation. We can also notice the absence of THz generation which we attribute to a limited broadening of the FWM visible mode.\\ 

Clearing up the angular spectral dynamics, Figs. \ref{Fig12}(b,c,d) detail the corresponding zones of interest where such secondary radiations develop near the main pump and FWM modes. The spectral extent in $k_r/k_0$ becomes increased by an averaged factor $\sim 3$ compared with the meter-range filamentation setup [see Figs. \ref{Fig12}(b,c,d) and Figs. \ref{Fig10}(d,e,f)]. Note that these angular spectra are expressed in frequency here. Typically, for the $S3$ mode (FW+SW), Fig. \ref{Fig12}(e) shows the angle-wavelength distribution that spans in angle up to 25$^{\circ}$ as function of $\lambda$, whereas the same emission angle drops to 6-8$^{\circ}$ in Figs. \ref{Fig12}(f) showing the equivalent pattern in the ELI filamentation experiment. Despite the difference in the input laser energies between the ELI setup and the Vilnius long-lensed setup, both operate in loosely focused geometry ($f \sim 100$ cm) and these behaviors quantitatively agree with the experimental angular patterns reported in Figs. \ref{Fig6}(a) and \ref{Fig5}(d), which highlight a factor $\sim 3.3$ difference in their conical emissions between a loosely and tighter focused geometry.\\

In summary, over extended propagation ranges the early Kerr stage amplifies the FWM modes and contributes to broaden enough the visible spectrum, which appears as a sine-qua-non condition to generate secondary emissions in the plasma regime. The plasma-induced THz band is mainly produced by the ionizing SW and broadened visible radiation. Most of these features can be retrieved in focused propagation regime, provided that the previous requirements on the broadening of the first FWM radiated modes be fulfilled.

\section{Conclusion}
\label{sec4}

We have investigated the generation mechanisms of mid-IR and visible radiation in air, not only by the classical Kerr-driven four-wave mixing, but also through the plasma response associated with Brunel emission process \cite {Brunel1990,David2025}. Experimental emphasis has been given to mid-IR emissions in gases, for which it was shown that four-wave mixing can be phase matched within a wide range of generated mid-IR wavelengths. We also reported and explained the emergence of weaker, secondary radiations.\\

In summary, we have analyzed, both experimentally and theoretically, the tunable mid-IR radiation generation in air by focused femtosecond laser pulses with two nonharmonic frequencies. Our presented proof-of-principle experiment showed that by using air filamentation one can produce mid-IR beams within a wide wavelength range, from 3 to 8 $\mu$m. In addition we addressed the possibility to generate, from two-color filaments, lower-amplitude secondary radiations centered around elementary combinations (difference, sum or their multiple) of the two laser pump frequencies, besides the mid-IR and visible radiations classically associated with four-wave mixing. By disentangling the Kerr and plasma regimes along the two-color filamentation range in LC and UPPE computations, we demonstrated that the early Kerr stage is necessary to amplify the FWM modes and it dominates in the conversion process for pump energies limited to a few mJ. However, a broad enough visible radiation and an efficient plasma response are required to generate secondary emissions, which, thereby, emerge as ''plasma spectral markers''. Plasma-induced THz radiation can moreover occur from the coupling of the ionizing SW pulse and broadened visible radiation near the pump nonlinear focus.\\

We believe that the systematic detection of such ''plasma markers'' in further experiments may help in discriminating between emissions triggered by optical nonlinearities and those created by the plasma itself.

\section*{Acknowledgement} 
This research has been carried out in the framework of the “Universities’ Excellence Initiative” programme by the Ministry of Education, Science and Sports of the Republic of Lithuania under the agreement with the
Research Council of Lithuania (project No. S-A-UEI-23-6). This work also benefited from access to the computing resources of the “CALI 3” cluster. This cluster is operated and hosted by the University of Limoges. It is part of the HPC network in the Nouvelle-Aquitaine Region, financed by the State and the Region.

\appendix

\section{Analytical justification of secondary radiations triggered by the visible FWM mode}
\label{sec:AppA}

We here justify from a simple analytical approach why triggering secondary radiations need the action of the visible component being even detuned to some extent. For doing so, we consider an MPI-type response $W(t) \propto |E(t)|^{2K},\,K \geq 1$, as representative of both plasma and Kerr nonlinearities [for a Kerr response no integration in time of the nonlinear function $W(t)$ is required], in which we insert the three plane wave field Eq. (\ref{E2}) with constant amplitudes $E_j$ ($j=1,2,3$), relative phases $\phi$ and $\theta$, and a visible frequency $\omega_3 = 2\pi \nu_{3}$ detuned by $\delta \omega_3 = 2 \pi \delta \nu_{3}$. We here ignore any spectral shift in the pump frequencies. Using $\cos{x} = (\exp(ix) + \exp(-ix))/2$ and double Newton binomial formulas, the MPI rate can be expanded as 
\begin{equation}
    \label{E1}
    W(t) = \frac{W_0}{2^{2K}} \sum_{p=0}^{2K}\sum_{k=0}^{2K-p}\binom{2K}{p} \binom{2K-p}{k} E_1^{2K-p-k}E_2^kE_3^p \sum_{q_1=0}^{2K-p-k}\sum_{q_2=0}^{k}\sum_{q_3=0}^{p} \binom{2K-p-k}{q_1} \binom{k}{q_2} \binom{p}{q_3} \exp(i \chi_p t + i \Psi),
\end{equation}
where
\begin{align}
\label{chipsi}
\chi_p & = (2K-p-k-2q_1) \omega_1 + (k-2 q_2)\omega_2 + (p-2q_3)(\omega_3 + \delta \omega_3), \\ \nonumber
\Psi & = (k-2q_2) \phi + (p-2 q_3) \theta.
\end{align}
In the above expressions, $K,p,k,q_1,q_2$ and $q_3$ are all positive integers satisfying 
\begin{equation}
\label{cond}
0 \leq q_1 \leq 2K-p-k,\,\,\,0 \leq q_2 \leq k \leq 2K-p, \,\,\, 0 \leq q_3 \leq p \leq 2K.
\end{equation}
The electron density proceeds from $N_e(t) \propto W(t)/(i \chi_p)$ and we a priori assume $\chi_p \neq 0$. Inserting the expression of the visible frequency $\omega_3 = 2\omega_1 - \omega_2$ into Eq. (\ref{chipsi}) leads to 
\begin{equation}
    \label{A10}
    \chi_p = [2K+p-k-2(q_1+2q_3)] \omega_1 + [k-p+2(q_3-q_2)]\omega_2 + (p-2q_3)\delta \omega_3.
\end{equation}
To generate new frequency combinations such as $\Omega = (\omega_1\pm\omega_2), 2(\omega_1\pm\omega_2)$ etc. by multiplying $N_e(t)$ with the optical field Eq. (\ref{E2}), we need an MPI spectrum already containing the frequencies $2\omega_1 \pm \omega_2$. In the absence of visible detuning $(\delta \omega_3 = 0)$, this simple constraint leads to the relationship
\begin{equation}
\label{imp}
q_1+q_2+q_3 = K - 1 \pm 1/2,
\end{equation}
which is impossible to satisfy and thus invites us to consider a non zero $\delta \omega_3$.\\

Let us for instance focus on the main hypothetical plasma mode $\chi_p = 2 \omega_1 - \omega_2$. We can verify that secondary radiations are permitted whenever, e.g., $p-k = 1 - 2(q_2-q_3)$ and
\begin{equation}
    \label{cond2}
    \frac{\delta \omega_3}{\omega_3} = \frac{1+q_1+q_2+q_3 - (K+1/2)}{(p-2q_3)}.
\end{equation}
Selecting a priori the integer relationships $q_1 = K$, $q_2 = q_3 = 0$ and $p=2$, the previous expression reduces to
\begin{equation}
    \label{cond3}
    \frac{\delta \omega_3}{\omega_3} = \frac{1}{4},
\end{equation}
which can easily be satisfied by photon numbers $K \geq 2(p-1) = 2$. This result agrees with the spectral distortions revealed by Fig. \ref{Fig7}(f). Note that when one relies on the Kerr effect alone ($K = 1$), secondary radiations are still possible with, e.g., $q_1 = p = 1$ ($q_2=q_3=0$) but for larger detunings $\delta \omega_3/\omega_3 = 1/2$. They can of course be triggered for smaller bandwidths and alternative combinations of integers, e.g., $\delta \omega_3/\omega_3 = 10\%$ with $q_1 = K > p = 5,\, q_2=q_3 =0$. Similar evaluations can moreover be derived for negative detunings, $\delta \omega_3 < 0$ [see Fig. \ref{Fig7}(e)], by repeating the above analytical steps for $\chi_p = \omega_2-2 \omega_1$.\\

Note that the same reasoning could have been applied to a two-color plane wave arrangement $(E_3=0)$, leading to the reduced exponential contributions $\chi_p^{2C},\,\Psi^{2C}$:
\begin{align}
\label{chipsi2C}
\chi_p^{2C} & = (2K-p-2q_1)\omega_1 + (p-2q_2)\omega_2, \\ \nonumber
\Psi^{2C} & = (p-2 q_2) \phi, \\ \nonumber
0 & \leq q_1 \leq 2K-p,\,\,\, 0 \leq q_2 \leq p \leq 2K.
\end{align}
To trigger the FWM frequencies $\Omega = 2\omega_i - \omega_j$ ($i\neq j=1,2$), we need a plasma spectrum promoting the frequency combinations $\chi_p^{2C} = \omega_i-\omega_j$. This constraint leads to the conditions $K=q_1+q_2,\,\,\,p=2q_2-1$ and is compatible with photon numbers $K \geq q_1+q_2$, which includes the classical Kerr-driven FWM emissions whenever $q_1$ or $q_2$ reduces to unity or zero. Similarly, creating secondary radiations from two colors only with no detuning in the second wave would lead to the constraint $q_1+q_2=K - 1/2$, which is impossible. Admitting a detuning $\delta \omega_2$ in the second wave would, however, require to fulfill the condition $\delta \omega_2 > \omega_2$, which is physically a nonsense, thereby confirming the need of a third color.\\

\bibliography{references.bib} 

\end{document}